\documentclass{aastex61} 
\usepackage{float}

\received{\today}

\shorttitle{M-Dwarf UV \& Prebiotic Chemistry}

\begin{document}
\title{The Surface UV Environment on Planets Orbiting M-Dwarfs: Implications for Prebiotic Chemistry \& Need for Experimental Follow-Up}

\correspondingauthor{Sukrit Ranjan}
\email{sranjan@cfa.harvard.edu}

\author[0000-0002-5147-9053]{Sukrit Ranjan}
\affiliation{Harvard-Smithsonian Center for Astrophysics, Cambridge, MA 02138, USA}
\affiliation{Harvard Origins of Life Initiative}

\author{Robin Wordsworth}
\affiliation{Harvard Origins of Life Initiative}
\affiliation{Harvard Paulson School of Engineering and Applied Sciences, Harvard University, Cambridge, MA 02140, USA}
\affiliation{Department of Earth and Planetary Sciences, Harvard University, Cambridge, MA 02140, USA}

\author{Dimitar D. Sasselov}
\affiliation{Harvard-Smithsonian Center for Astrophysics, Cambridge, MA 02138, USA}
\affiliation{Harvard Origins of Life Initiative}
\affiliation{Simons Collaboration on the Origin of Life}

\begin{abstract} 
Potentially-habitable planets orbiting M-dwarfs are of intense astrobiological interest because they are the only rocky worlds accessible to biosignature search over the next 10+ years due to a confluence of observational effects. Simultaneously, recent experimental and theoretical work suggests that UV light may have played a key role in the origin of life on Earth, and especially the origin of RNA. Characterizing the UV environment on M-dwarfs planets is important to understanding whether life as we know it could emerge on such worlds. In this work, we couple radiative transfer models to observed M-dwarf spectra to determine the UV environment on prebiotic Earth-analog planets orbiting M-dwarfs. We calculate dose rates to quantify the impact of different host stars on prebiotically-important photoprocesses. We find that M-dwarf planets have access to ~100-1000 times less bioactive UV fluence than the young Earth. It is unclear whether UV-sensitive prebiotic chemistry that may have been important to abiogenesis, such as the only known prebiotically plausible pathways for pyrimidine ribonucleotide synthesis, could function on M-dwarf planets. This uncertainty affects objects like the recently-discovered habitable-zone planets orbiting Proxima Centauri, TRAPPIST-1, and LHS 1140. Laboratory studies of the sensitivity of putative prebiotic pathways to irradiation level are required to resolve this uncertainty. If steady-state M-dwarf UV output is insufficient to power these pathways, transient elevated UV irradiation due to flares may suffice; laboratory studies can constrain this possibility as well. 
\end{abstract}




\section{Introduction\label{sec:intro}}
Planets orbiting M-dwarfs are the most compelling targets for the search for life beyond the solar system. M-dwarfs are the most common type of star in the Galaxy, and exoplanet population studies suggest potentially habitable\footnote{i.e. Earth-sized planets receiving instellation compatible with surface liquid water} planets orbiting these stars are common \citep{Dressing2015}. Indeed, Proxima Centauri, the closest star to our Solar System and an M-dwarf, has recently been shown to host a potentially habitable world \citep{Anglada-Escude2016}, the late M-dwarf TRAPPIST-1 to host 3 \citep{Gillon2017}, and the nearby M-dwarf LHS 1140 to host another, well-suited for atmospheric characterization observations \citep{Dittmann2017}. Perhaps most importantly, due to a confluence of observational effects, M-dwarf terrestrial planets (and \emph{only} M-dwarf terrestrial planets) will be accessible to atmospheric characterization and hence biosignature search with the flagship telescopes\footnote{e.g. the James Webb Space Telescope, JWST, and the Extremely Large Telescopes, ELTs} due to come online over the next decade  \citep{Seager2014, Rodler2014, Batalha2015, Cowan2015}. Such observations will require tens to hundreds of hours of observation time on the best telescopes in the world, a very large investment. Consequently, it is crucial to understand whether life can emerge and endure on such worlds. 

Extensive work has been done on M-dwarf planet habitability, i.e., whether life as we know it could endure on these worlds; see, e.g.,  \citet{Tarter2007}, \citet{Scalo2007}, and \citet{Shields2016} for reviews of work on this topic. However, far fewer investigations have been conducted as to the favorability of M-dwarf planets for abiogenesis (the origin of life), i.e., whether life as we know it could emerge on these worlds. Investigations of the ease with which life can originate on different worlds are relevant to prioritization of targets for biosignature search, and for interpretation of putative biosignatures detected from such objects (e.g., \citealt{Catling2017}). In part, the paucity of work on this question is due to limitations in our understanding of the origin of life. For example, much of the work on exoplanet habitability has been motivated by the biological fact that life as we know it today requires liquid water, and has focused on understanding the availability of this key requirement for life in different planetary environments (e.g., \citealt{Kasting1993habzone}). Such certainty does not exist in our understanding of the origin of life on Earth, making it challenging to compare the clemency of planetary environments for abiogenesis scenarios.

Recent advances in prebiotic chemistry (chemistry relevant to the origin of life) are changing this situation. The last decade has seen breakthroughs in long-standing problems in prebiotic chemistry, perhaps most remarkably in the discovery of plausible mechanisms for the prebiotic synthesis of activated pyrimidine ribonucleotides \citep{Powner2009, Xu2016}, the selective synthesis of short (2- and 3-carbon) sugars \citep{Ritson2012}, and a reaction network generating precursors for each of the four fundamental classes of biomolecule (lipids, amino acids, carbohydrates, and nucleotides) \citep{Patel2015}. These advances represent major progress towards the solution of critical problems in prebiotic chemistry, including the asphaltization problem in sugar synthesis \citep{Benner2012} and the N-glcosylation of ribose with the nucleobases to make ribonucleosides and ribonucleotides\footnote{The monomers of RNA} \citep{Sponer2016}. These problems are decades-old, and their solution is required for the abiotic origin of RNA, and hence the origin of life in the RNA world hypothesis \citep{Gilbert1986}. 

UV light plays a key role in these recently proposed prebiotic pathways. While UV light can destroy nascent biomolecules \citep{Sagan1973}, it can also power synthetic prebiotic photochemistry.  In addition to the pathways discussed above, UV light is invoked in prebiotic chemistry as diverse as the origin of chirality \citep{Rosenberg2008}, the synthesis of amino acid precursors \citep{Sarker2013}, and the polymerization of RNA \citep{Mulkidjanian2003}. Measurements of nucleobase photostability suggest that the biogenic nucleobases (the informational components of the RNA and DNA monomers) are exceptionally stable to UV irradiation compared to structurally similar molecules with comparable thermal properties, suggesting they evolved in a UV-rich environment \citep{Rios2013, Beckstead2016, Pollum2016}. This scenario is consistent with our understanding of conditions on prebiotic Earth: UV light is thought to have been abundant on young Earth due to the absence of a biogenic ozone layer \citep{Cockell2000UVhist, Cockell2000oceans, Ranjan2016}. Prebiotic photochemistry interacts with UV radiation in ways that are sensitive to its spectral shape and overall intensity \citep{Ranjan2016}. Consequently, it is important to constrain the UV environment on the surface of planets orbiting M-dwarfs, to understand if UV-sensitive prebiotic chemistry pathways that could have lead to the origin of life on Earth could function on such worlds\footnote{We clarify that availability of adequate UV light is a necessary but insufficient requirement for UV-dependent prebiotic chemistry. Here, we focus on the availability of UV light, but other work is required to constrain whether other requirements for putative prebiotic chemistry are available on M-dwarf planets. An example is the work of \citet{Nava-Sedeno2016}, which suggests that reducing compounds like CO and HCN, which are invoked in many prebiotic chemistries, may be abundant on planets orbiting M-dwarfs.}. This represents a new criterion for planetary inhabitability, motivated by specific empirical advances in prebiotic syntheses and by an emergent theoretical understanding of the likely importance of high-energy radiation in origin-of-life scenarios \citep{Pascal2012}. 

In this work, we use a two-stream multilayer radiative transfer model to constrain the surface UV environment on planets analogous to prebiotic Earth orbiting M-dwarf stars. Our model includes the effects of absorption and multiple scattering from the surface and from atmospheric gases, and uses recently-measured high-quality UV observations of M-dwarf stars (e.g., \citealt{France2016}) to provide realistic top-of-atmosphere (TOA) stellar irradiation spectra. We convolve the calculated surface radiance spectra against action spectra corresponding to two fundamental, simple photochemical reactions (one useful to prebiotic chemistry, and one detrimental) that may have been important during the era of abiogenesis, integrate the results to compute measures of prebiotically-important reaction rates, and discuss the implications for the emergence of life on planets orbiting M-dwarfs. Our work suggests the need for specific experimental tests that must be done to determine whether the UV-sensitive prebiotic chemistry that may have powered the origin of life on Earth could function on planets orbiting M-dwarf stars.

\section{Background: Previous Studies of M-dwarf Planet UV}
Most habitability studies of M-dwarfs treat UV radiation as a negative for habitability \citep{Heath1999, Tarter2007, Lammer2009, Shields2016, Meadows2016}, motivated by the observation that UV radiation has deleterious effects on modern life \citep{Setlow1974}. Extensive work has been done to explore mechanisms that might protect surface life from the UV output of M-dwarfs in quiescence and in flare, such as ozone layers \citep{Segura2005, Segura2010, Rugheimer2015}, oceans \citep{Kiang2007}, and bioflourescence \citep{Omalley-James2016}. Overall, these works show that UV irradiance at the surface of modern-Earth analog M-dwarf planets should be suppressed to levels below those of modern Earth itself, due to lower M-dwarf near-UV (NUV) output and due to favorable ozone-generating atmospheric photochemistry. Similarly, the UV levels on M-dwarf planets analogous to young Earth (i.e. with anoxic atmospheres) should be lower than on young Earth itself, due to lower M-dwarf NUV emission. Hence, these works in aggregate conclude that UV radiation environment on terrestrial planets orbiting M-dwarf stars should be clement for life, with potential caveats for very active stars\footnote{These arguments presume that planetary atmospheres can be retained despite processes like XUV-powered atmospheric escape, which has been demonstrated for super-Earths ($M_{planet}\geq 6 M_\Earth$, \citealt{Tian2009b}). However, escape calculations for young Mars ($M=0.1 M_\Earth$)suggest its atmosphere would have been unstable to escape powered by higher XUV emission from the young Sun \citep{Tian2009a}. It is therefore unclear whether the atmospheres Earth-mass planets orbiting M-dwarfs should be stable to escape, particularly in light of M-dwarf's enhanced fractional XUV emission relative to solar-type stars. Further modelling is required to constrain this possibility.}. In this work, we consider in addition the possible positive roles of UV irradiation, motivated by recent experimental advances in prebiotic chemistry that require UV light.

Some previous workers have considered possible positive roles for UV radiation for early life. \citet{Scalo2007} hypothesize that highly variable M-dwarf UV emission could drive variations in mutation rates on orbiting planets, which might enhance the rate of evolution. \citet{Buccino2007} argue based on the ``Principle of Mediocrity" \footnote{The hypothesis that Earth's properties should be typical of inhabited planets.} that habitable planets should receive stellar irradiance similar to Archaean Earth in order to power potential prebiotic chemistry, and use it to suggest life cannot arise on planets orbiting inactive, low-UV M-dwarfs because in order to receive Earthlike UV instellation, planets will need to orbit within the inner edge of the habitable zone. \citet{Buccino2007} suggest that moderately active M-dwarfs may consequently be better candidates for habitability due to enhanced UV output during flares. These works are abstract in their arguments, and neither links their discussion of UV environment to specific prebiotic photochemistry. Our work is differentiated from these works in coupling the M-dwarf UV environment to specific prebiotic photoreactions through their action spectra. Our work is also differentiated from \citet{Buccino2007} in computing spectrally resolved radiation environments, considering the role of atmospheric attenuation, and using a larger, higher-quality sample of M-dwarf UV radiation fields.

\section{Methods}
In this section, we describe our methods. In brief, we calculated the attenuation of empirically measured M-dwarf UV emission by the atmosphere to compute the spectral surface radiance, and coupled the spectral radiance to prebiotically relevant action spectra to evaluate the implications for prebiotic chemistry. We chose a two-stream approach to radiative transfer to correctly account for the role of multiple scattering in atmospheric attenuation \citep{Ranjan2017b}, thought our results are ultimately insensitive to inclusion of this effect. All code associated with this project is available for validation and extension at \url{https://github.com/sukritranjan/ranjanwordsworthsasselov2017b} (Zenodo archive: \citealt{ZenodoRelease}).

\subsection{Radiative Transfer}

Our radiative transfer formalism is described in detail in \citet{Ranjan2017a}. Briefly, we partition the atmosphere into 64 1-km homogenous layers and use a two-stream formalism with Gaussian (single) quadrature closure to compute propagation of UV light through the atmosphere \citep{Toon1989}. Two-stream radiative transfer is monochromatic; we partition our spectra into wavelength bins of 4 nm width\footnote{chosen to avoid negative fluxes in low-SNR input spectra, see Section~\ref{sec:methods_inputfluxes}.}, and integrate all optical parameters over these bins. For numerical stability, we assign a ceiling on the per-layer single-scattering albedo $\omega_0$ of $(1-1^{-12})$. We take the surface albedo to be 0.2, a representative value for rocky planets and consistent with past 1D modelling for Earth and Mars \citep{Kasting1991, Segura2003, Wordsworth2015}. We take the solar zenith angle (SZA) to be $48.2^\circ$, corresponding to the insolation-weighted global mean value \citep{Cronin2014}. These parameter value correspond to global mean conditions; variations in surface albedo and solar zenith angle can generally be expected to drive changes in band-averaged fluence of approximately 1 order of magnitude \citep{Ranjan2017a}. As in \citet{Ranjan2017a} and \citet{Ranjan2017b}, the fundamental quantity our code calculates is the surface radiance, i.e. the integral of the intensity field at the planet surface for elevations $>0$. This is the relevant radiative quantity for calculating photoreaction rates at planet surfaces, and in the two-stream formalism can be calculated as $$I_{surf}=F^{\downarrow}_{N}/\mu_1+F^{dir}_N/\mu_0,$$ where $I_{surf}$ is the surface radiance, $F^{\downarrow}_{N}$ is the downward diffuse flux at the planet surface, $F^{dir}_N$ is the direct flux at the planet surface, $\mu_0=\cos(SZA)$ is the cosine of the solar zenith angle, and $\mu_1=1/\sqrt{3}$ for Gaussian quadrature \citep{Toon1989}.

\subsection{Atmospheric Model}
Following \citet{Rugheimer2015}, we take the planetary atmosphere to be cloud-free, with a surface pressure of 1 bar composed of 0.9 bar N$_2$ and 0.1 bar CO$_2$. Our results are insensitive to these assumptions on atmospheric state, because CO$_2$ absorption ($<204$ nm) saturates for $p_{CO_{2}}>0.072$ bar\footnote{Such high levels of CO$_2$ are expected for abiotic Earth-analogs because CO$_2$ is emitted in bulk from volcanos for planets with Earthlike mantle oxidation states.}. The absorption of most major atmospheric gases (e.g., N$_2$, H$_2$O, CH$_4$) is degenerate with this CO$_2$ absorption, meaning that surface UV is insensitive to their abundances \citep{Ranjan2017a}. Trace gases whose UV absorption is nondegenerate with CO$_2$ (e.g., O$_3$, SO$_2$) do not build up to levels high enough to affect surface UV for atmospheric boundary conditions corresponding to the modern abiotic Earth, according to the photochemical calculations of \citet{Rugheimer2015}\footnote{For atmospheric boundary conditions not corresponding to the modern abiotic Earth, it is possible for elevated levels of O$_3$ and O$_2$ to build up on M-dwarf planets, e.g., \citet{Harman2015}. Section~\ref{sec:disc_1} discusses this phenomenon and its implications in more detail.}. Similarly, the absorption of both CO$_2$ and H$_2$O clouds are degenerate with gaseous CO$_2$ absorption,  and for Earthlike global mean cloud optical depths of 4-10 \citep{Stubenrauch2013}, the surface UV environment is insensitive to the presence of clouds \citep{Ranjan2017b}. Finally, because the thermal emission of the atmosphere is negligible at UV wavelengths, the UV surface radiance is insensitive to the atmospheric temperature/pressure profile. Consequently, we approximate the atmosphere by a simple exponential model, with $T=T_{surf}=288K$ throughout. We experimented with a more realistic atmospheric model with dry adiabatic evolution in the troposphere and an isothermal stratosphere, and obtained conclusions identical to the exponential atmosphere; consequently, we elected to use the simpler exponential model in this work.

\subsection{Stellar Fluxes\label{sec:methods_inputfluxes}}
We used empirically measured UV spectra of M-dwarfs as top-of-atmosphere (TOA) fluxes to input into our radiative transfer model. Comparatively few such measurements are available, due to low M-dwarf luminosity and high telluric opacity in the UV. We relied primarily on spectra collected by the MUSCLES project, which obtained high-quality spectra of 7 M-dwarfs with the Hubble Space Telescope \citep{France2016, Loyd2016} and combined them with measurements from other instruments and stellar models to aggregate broadband emission spectra; we use all M-dwarf stars in their sample\footnote{Accessed via \url{https://archive.stsci.edu/prepds/muscles/}, 2016 December 16.}. The MUSCLES team also aggregated HST and XMM-Newton measurements and a PHOENIX stellar model to create a spectrum for Proxima Centauri, which we also use. \citet{Segura2005} aggregated IUE and HST observations to compile a UV spectrum of AD Leo in quiescence; we used this spectrum as well\footnote{Accessed via \url {http://vpl.astro.washington.edu/spectra/stellar/mstar.htm}, 2016 December 16.}.  Because of their low luminosities, no spectra of M-dwarfs of stellar type later than M5 are available in the NUV regime relevant to surficial prebiotic chemistry\footnote{Though measurements of late M-dwarf emission at shorter wavelengths are available; see, for example, the detection of Lyman-$\alpha$ radiation from the M8 star TRAPPIST-1 by \citet{Bourrier2017}. We may hope for NUV spectra of such objects to be one day accessible, perhaps through extended observations with space-based instruments.}. To obtain coverage of late-type M-dwarfs, we use the M8 ``active" model of \citet{Rugheimer2015}, which is formed by scaling a spectrum of AD Leo using emission lines and concatenating it to a PHOENIX model. This may be taken to correspond to the spectrum of a highly active M8 star. 

M-dwarfs are known for their frequent and energetic flares \citep{Osten2016}. Few spectral measurements of M-dwarf flares in the prebiotically relevant 150-300 nm regime are available. The exception is AD Leo; the Great Flare of 1985 on this star was measured by space- and ground-based instruments, and \citet{Segura2010} have synthesized these measurements to compile spectra of this flare at different time points from 100-444 nm. We use these flare spectra of AD Leo to understand the impact of M-dwarf flares on the surface UV environment and on prebiotic chemistry. We note that the Great Flare on AD Leo may not be representative of all M-dwarf flares, as it was an exceptionally energetic flare (U-band emission of $10^{33.8}$ erg, \citealt{Hawley1991}) on one of the most active known M-dwarfs\footnote{See, e.g., \citet{Kowalski2013}}. However, it is the only broad-coverage (spectral and temporal), spectrally-resolved measurement of an M-dwarf flare we are aware of, and hence remains the focus of our investigation. Results derived from this flare may be interpreted as a limiting case of the impact of M-dwarf flares on the surface UV environment. 

We bin all data to a resolution of 4 nm, to eliminate negative fluxes. Such a coarse resolution is acceptable when working with biological action spectra, which have spectral features with widths on the order of $\gtrsim10$ nm (see, e.g., \citet{Setlow1974}, \citealt{Ronto2003}, \citealt{Cnossen2007}). Following \citet{Segura2005}, we scale all M-dwarf emission spectra to a distance $a$ such that the flux at distance $a$ is equal to the modern Solar constant, including the factor of 0.9 correction for the redshifted SEDs of M-dwarfs: $$a=\sqrt{\frac{L/L_\Sun}{0.9}} * 1 AU$$ Table~\ref{tbl:methods_mdwarfproperties} provides these distances and also summarizes other relevant properties of the M-dwarfs in our sample.  

To establish a basis of comparison between M-dwarfs and Sunlike stars, we use the model of \citet{Claire2012} to calculate the spectra of the young Sun at 3.9 Ga. We choose this age because it is consistent with available evidence for the origin of life on Earth (see, e.g.,  \citealt{Ranjan2016}, Appendix A); our results are insensitive to the choice of solar age, and are unaffected for solar ages from 3.5-4.1 Ga.

\begin{table}[H]
\begin{center}
\caption{Stars Used in This Study \& Associated Properties. \label{tbl:methods_mdwarfproperties}}
\begin{tabular}{p{1.8 cm}p{1.5 cm}p{1.8 cm}p{1.2 cm}p{1.2 cm}p{4 cm}p{3 cm}}
\tableline\tableline
Star & $T_{eff} (K) $ & Spectral Class & $d$ (pc) & $a (AU)$  & Reference & Note\\
GJ 1214 & 2953 & M4.5 & 14.6 & 0.064 & 1, and sources therein &\\
Proxima Centauri & 3042 & M5.5 & 1.3 & 0.042 & 2, 3, and sources therein &Flare star\\
GJ 876 & 3062 & M5 & 4.7 & 0.12 & 1, and sources therein &\\
GJ 436 & 3281 & M3.5 & 10.1 & 0.17 & 1, and sources therein &\\
GJ 581 & 3295 & M5 & 6.2 & 0.11 & 1, and sources therein &\\
GJ 667C & 3327 & M1.5 & 6.8 & 0.11 & 1, and sources therein &\\
AD Leo & 3400 & M3.5 & 4.9 & 0.16  & 4, 5, 6 and sources therein & Extremely active star\\
GJ 176 & 3416 & M2.5 & 9.3 & 0.19 & 1,  and sources therein &\\
GJ 832 & 3816 & M1.5 & 5.0 & 0.14 & 1, and sources therein &\\
\tableline
\tableline
\end{tabular}
\end{center}

\tablecomments{\textbf{References}. (1) \citet{Loyd2016}. (2) \citet{Anglada-Escude2016}. (3) \citet{Meadows2016}. (4) \citet{Shkolnik2009}. (5) \citet{Rojas-Ayala2012}. (6) \citet{Segura2005}.}
\end{table}

\subsection{Action Spectra and Calculation of Dose Rates\label{sec:doserates}}

To quantify the impact of different UV surface radiation environments on prebiotic photochemistry, we compute Biologically Effective Dose rates (BEDs), which measure the reaction rates of specific prebiotically important photoprocesses \citep{Cockell1999, Ronto2003, Rugheimer2015}. Our method is described in detail in \citet{Ranjan2017a} and \citet{Ranjan2017b}. Briefly, we compute $$D=(\int_{\lambda_{0}}^{\lambda_{1}} d\lambda \mathcal A(\lambda)I_{surf}(\lambda)). $$ $I_{surf}(\lambda)$ is the UV surface radiance, computed from our model. $\mathcal A(\lambda)$  corresponds to the action spectrum, which parametrizes the wavelength dependence of a given photoprocess; higher values of $\mathcal A$ mean that a higher fraction of the incident photons are being used in the photoprocess. $\lambda_0-\lambda_1$ is the wavelength range over which $\mathcal A(\lambda)$ and $I_{surf}(\lambda)$ are defined. Since $D$ is a relative measure of reaction rate, a normalization is required to assign a physical interpretation to its value. In this paper, we report $$\overline{D}=D/D_{\earth},$$ where $D_{\earth}$ is the dose rate on 3.9 Ga Earth. $\overline{D}>1$ means the reaction is proceeding faster than it would have on young Earth; $\overline{D}<1$, the reverse.

We use action spectra corresponding to simple, prebiotically relevant photoprocesses to measure the impact of UV light on nascent life. This differentiates our work from previous works \citep{Cockell2000UVhist, Cockell2002, Cnossen2007, Rugheimer2015} which used action spectra corresponding to DNA damage in modern organisms. Focus on DNA damage is inappropriate for prebiotic chemistry because 1) DNA is not thought to have been the primordial biomolecule, 2) modern organisms have evolved sophisticated methods to deal with environmental stress, including UV exposure, that would not have been available to the first life, and 3) this approach ignores the role of UV light as a eustressor\footnote{i.e. beneficial for} for abiogenesis. We consider two photochemical reactions: a stressor process and a eustressor process. We also compute the band-integrated NUV radiation, which is a pathway-independent measure of the abundance of prebiotically useful radiation. 

For our stressor process, we use the cleavage of the N-glycosidic bond in the RNA monomer uridine monophospate (UMP) by UV light, which irreversibly destroys this key biomolecule. We take the action spectrum as the product of the UMP absorption spectrum \citep{Voet1963} and the quantum yield curve. We assume a step function form to the quantum yield curve, with value $4.3\times10^{-3}$ for $\lambda\leq\lambda_0$ and $2.5\times10^{-5}$ for $\lambda>\lambda_0$, and we consider $\lambda_0$ values of 193, 254, and 230 nm, consistent with the measurements of \citet{Gurzadyan1994}. The absorption spectra of the other RNA monomers are structurally similar to UMP and the molecules share many photochemical properties\footnote{e.g, the quantum yield of N-glycosidic bond cleavage in adenosine monophospate (AMP) increases at short wavelengths like UMP's does \citep{Gurzadyan1994}}. Therefore, if a UV environment is destructive for UMP, it should be destructive for the other RNA monomers, and hence for abiogenesis in the RNA world hypothesis, as well. As shorthand, we refer to this photoprocess under the assumption that $\lambda_0=$X nm by UMP-X. 

For our eustressor process, we use the production of solvated electrons from the irradiation of a tricyanocuprate (CuCN$_3^{2-}$) complex, which was invoked by \citet{Ritson2012} in their pathway for the selective synthesis of the short sugars (glycolaldehyde, gylceraldehyde) that are key for the synthesis of RNA, and which are generally useful in a wide range of reductive prebiotic chemistry. We again take the action spectrum to be the product of the absorption spectrum \citep{Magnani2015, Ranjan2016} and the quantum yield curve. Following \citet{Ritson2012}'s hypothesis that photoionization of the cyanocuprate drives solvated electron production, we assume the QY to be characterized by a step function with value $0.06$ for $\lambda\leq\lambda_0$ and $0$ otherwise, and we consider $\lambda_0=254$ and 300 nm, consistent with the empirical constraints of \citet{Ritson2012}. Laboratory measurements of the spectral quantum yield of this process are forthcoming \citep{Todd2017a}; preliminary results suggest a step occurring at $\sim250$ nm. As shorthand, we refer to this photoprocess under the assumption that $\lambda_0=$Y nm by CuCN3-Y. 

\begin{figure}[H]
\centering
\includegraphics[width=10 cm, angle=0]{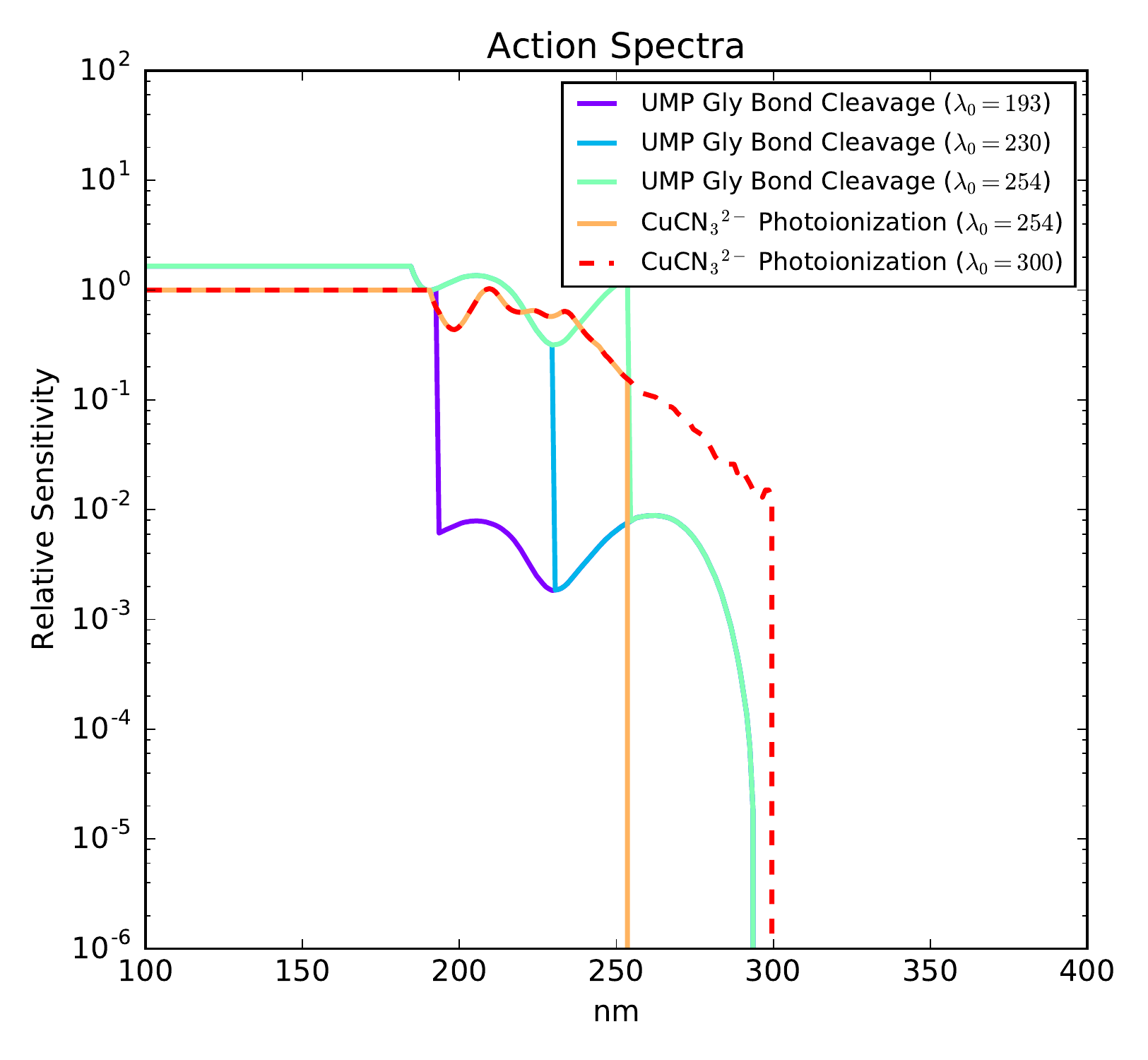}
\caption{Action spectra for photolysis of UMP$-\lambda_0$ and photoionization of CuCN3 $-\lambda_0$, assuming a step-function form to the QE for both processes with step at $\lambda_0$. The spectra are arbitrarily normalized to 1 at 190 nm. Data shortward of 184 nm for UMP$-\lambda_0$ and CuCN3$-\lambda_0$ are padded.\label{fig:actspec}}
\end{figure}

Action spectra typically encode information about relative, not absolute, reaction rates. Consequently, they are generally arbitrarily normalized to 1 at some wavelength (see, e.g., \citealt{Cockell1999} and \citealt{Rugheimer2015}). We normalize these spectra to 1 at 190 nm. We were unable to located absorption cross-section data for tricyanocuprate for $\lambda<190$ nm, nor for UMP for $\lambda<184$ nm; for wavelengths below these cutoffs, we padded the absorption spectra with the shortest wavelength data available. Since  $\lambda<190$ nm is shielded out by even modest amounts of CO$_2$ in the atmosphere, this padding only comes into play when considering nonexistent or tenuous atmospheres, as we briefly consider in Section~\ref{sec:disc}. Figure~\ref{fig:actspec} presents the action spectra.

The action spectra discussed above correspond to specific prebiotically relevant photochemical processes. In addition, we calculate the band-integrated surface radiance for $\lambda=200-300$ nm, which we term the "NUV radiance". This is based on emerging studies of the wavelength-dependence of a number of prebiotically important photoprocesses, which suggest that radiation in the 200-300 nm regime can be useful to prebiotic chemistry, while radiation at wavelengths $<200$ nm seems to be solely destructive \citep{Todd2017a}. We use the NUV radiance as a process-independent measure of the abundance of prebiotically useful UV radiation; the dose rate for the NUV radiance tracks those of the specific prebiotic photoprocesses, as expected for a generalized measure. 

\section{Results}

\subsection{Steady-State M-dwarf Emission}
\begin{figure}[H]
\centering
\includegraphics[width=10 cm, angle=0]{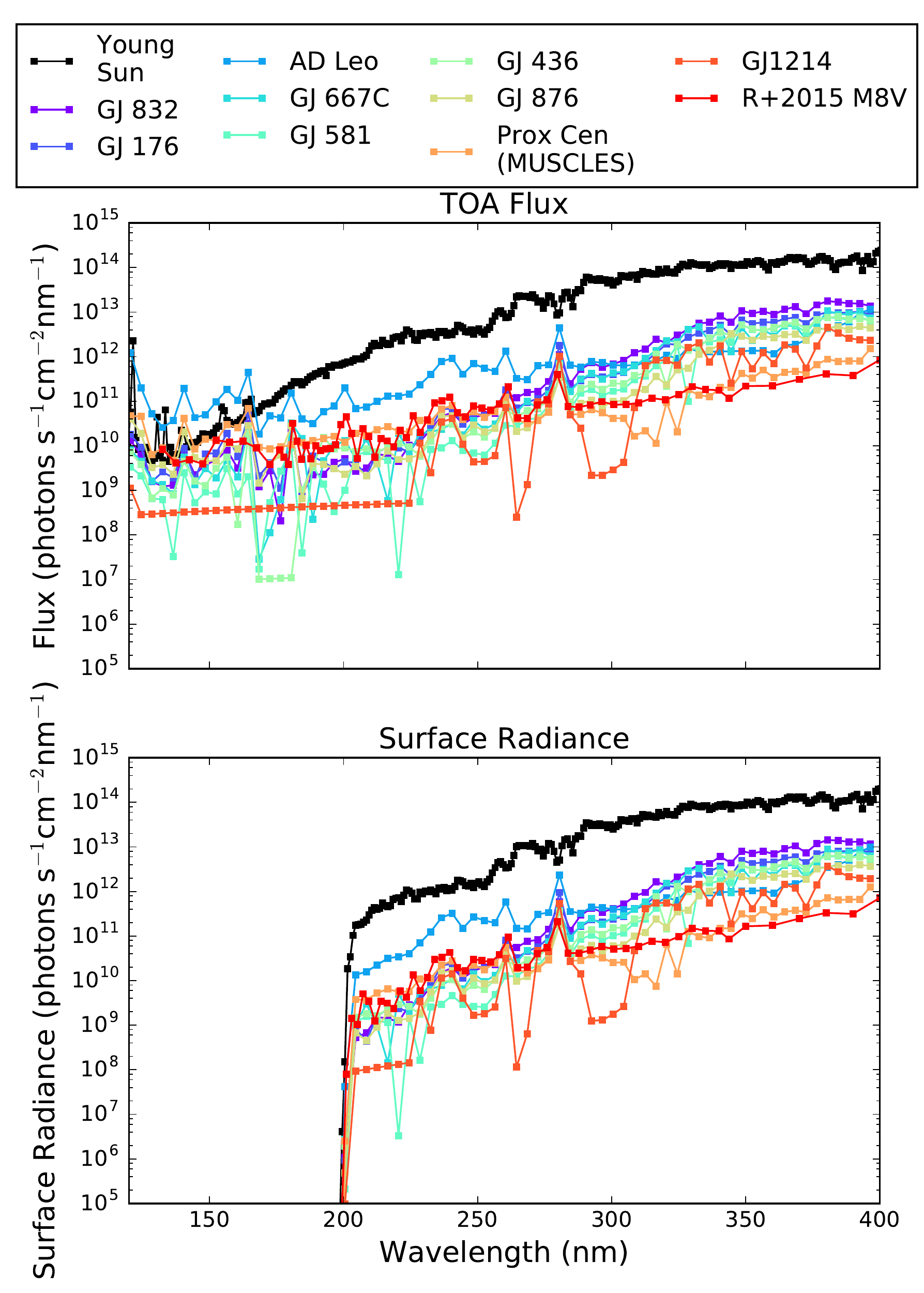}
\caption{TOA fluxes (top) and corresponding surface radiances as a function of wavelength for prebiotic Earth-analog planets with 1-bar N$_2$-CO$_2$ atmospheres orbiting a range of M-dwarfs. The young Sun and Earth case (3.9 Ga) is also shown for comparison. \label{fig:results_radiances_steadystate}}
\end{figure}

Figure~\ref{fig:results_radiances_steadystate} presents the TOA fluxes for the M-dwarfs in our study and the corresponding surface radiances on prebiotic Earth-analog planets shielded by a 1 bar atmosphere (0.9 bar N$_2$, 0.1 bar CO$_2$, matching \citealt{Rugheimer2015}). As noted by \citet{Rugheimer2015}, the surface conditions are defined by a cutoff at 204 nm imposed by atmospheric CO$_2$, and minimal attenuation at longer wavelengths. Notably, young-Earth-analog planets orbiting M-dwarf planets are exposed to far less UV radiation than those orbiting Sunlike stars, because of the cooler photosphere and hence lower NUV emission of M-dwarfs. At short wavelengths, M-dwarfs emit proportionately more radiation than the young Sun, but fluence at these wavelengths is robustly blocked by a range of atmospheric absorbers, including CO$_2$ and H$_2$O, which shield out $<200$ nm radiation. Consequently, M-dwarf planets, so long as they can retain their atmospheres, are low-UV environments.

\begin{figure}[H]
\centering
\includegraphics[width=10 cm, angle=0]{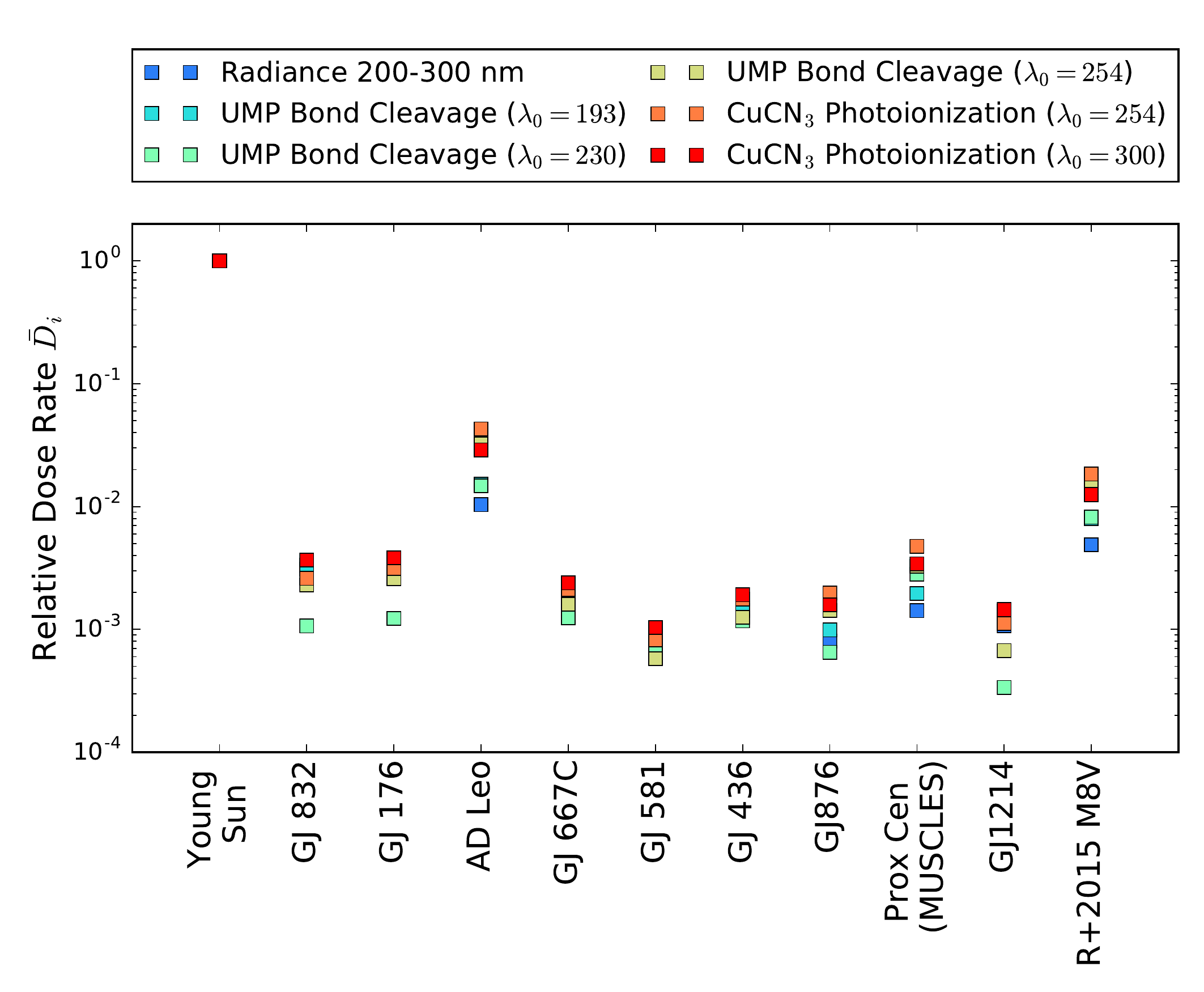}
\caption{UV dose rates $\overline{D}_i$ for UMP-X and CuCN3-Y at the surfaces of prebiotic Earth-analog planets with 1-bar N$_2$-CO$_2$ atmospheres orbiting a range of M-dwarfs. Dose rates corresponding to the 3.9 Ga Earth are also shown for comparison. \label{fig:results_doserates_steadystate}}
\end{figure}

We quantify this observation by computing the relative dose rates $\overline{D}$ for each of the stars in our sample. These dose rates are presented in Figure~\ref{fig:results_doserates_steadystate}. The stars are ordered by decreasing $T_{eff}$. We observe that with the exception of the exceptionally active star AD Leo and the \citet{Rugheimer2015} M8V "Active" model, which is a scaled AD Leo, all dose rates are suppressed by $>2$ orders of magnitude relative to the young Earth. Putative UV-dependent prebiotic chemistry will proceed at rates 2-4 orders of magnitude slower on planets orbiting non-active M-dwarfs compared to the young Earth. Active stars like AD Leo emit more UV radiation, so bioactive fluence on planets orbiting such active stars will only be suppressed by 1-2 orders of magnitude.

\begin{figure}[H]
\centering
\includegraphics[width=10 cm, angle=0]{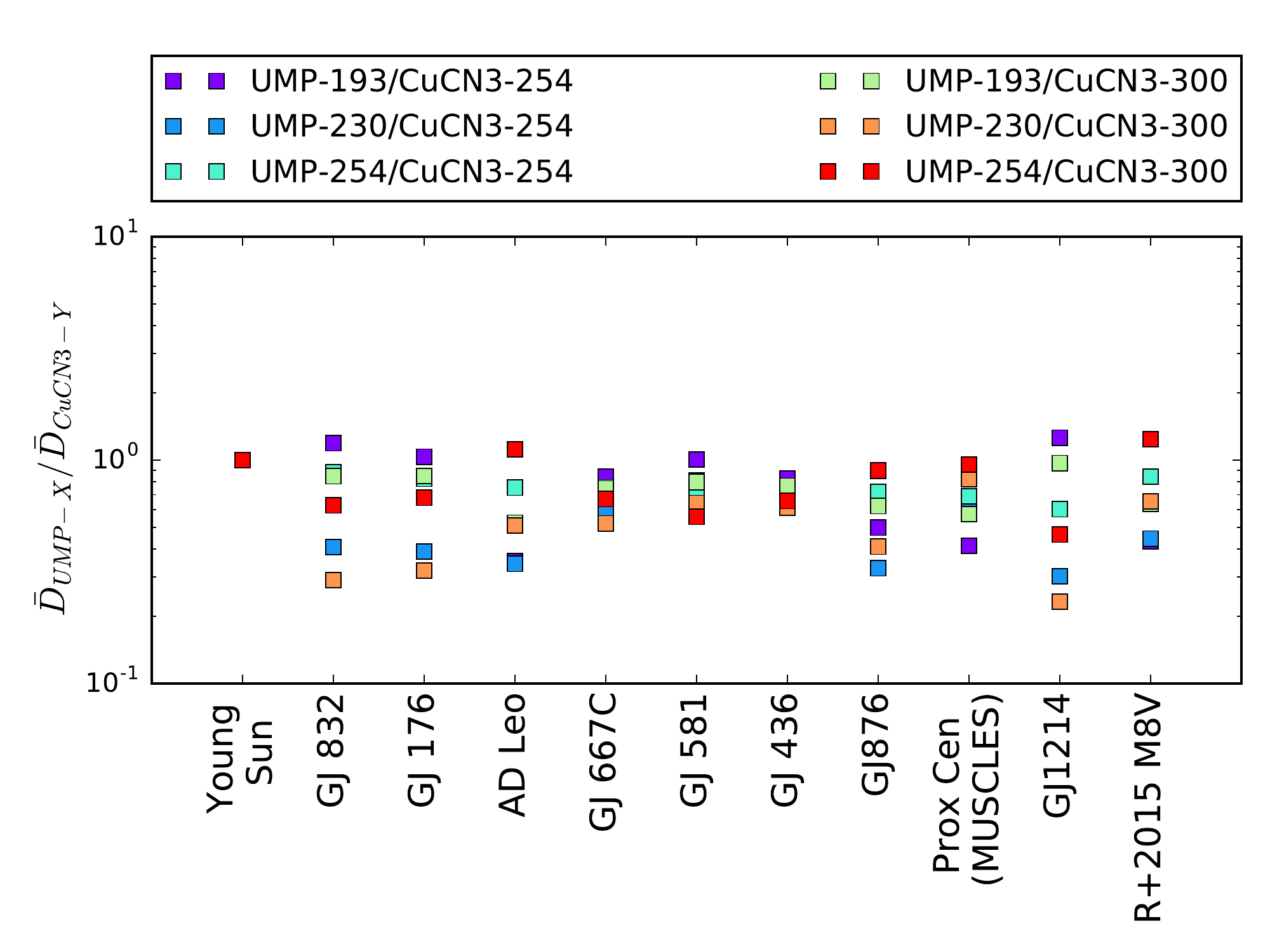}
\caption{Ratio of stressor dose rates UMP-X divided by eustressor dose rates UMP-Y for dusty CO$_2$-H$_2$O atmospheres, as a function of host star. Lower ratios imply a more favorable environment for abiogenesis as measured by these two photoprocesses. \label{fig:results_reldoserates_steadystate}}
\end{figure}

We considered the hypothesis that different host stars might affect the stressor and eustressor pathways in different ways. Specifically, we considered the possibility that the eustressor reaction rates might fall off slower or faster than the stressor reaction rates due to variations in the shape of the Spectral Energy Distributions (SEDs) between the different host stars, which would imply more or less favorable venues for abiogenesis, respectively. To test this hypothesis, we calculated $\overline{D}_{UMP-X}/\overline{D}_{CuCN3-Y}$ for all $X$ and $Y$. If this ratio is $<1$, the stressor pathway is disfavored over the eustressor pathway relative to Earth, and the environment compares favorably to Earth as a venue for abiogenesis; if this ratio is $>1$, the reverse is true. Figure~\ref{fig:results_reldoserates_steadystate} presents these calculations. In aggregate, the dose rate ratios does not vary much as a function of host star, remaining within an order of magnitude of unity for all stars in this study, and no clear aggregate trends are visible. We attribute this to the fact that the SEDs of the M-dwarf stars considered here are broadly similar in shape to the Sun's SED at the NUV wavelengths that makes it to the planet surface. We conclude that M-dwarf planets are comparable to early Earth in terms of how much their UV environments favor stressor processes over eustressor processes.

\subsection{M-dwarf Flares}
\begin{figure}[H]
\centering
\includegraphics[width=10 cm, angle=0]{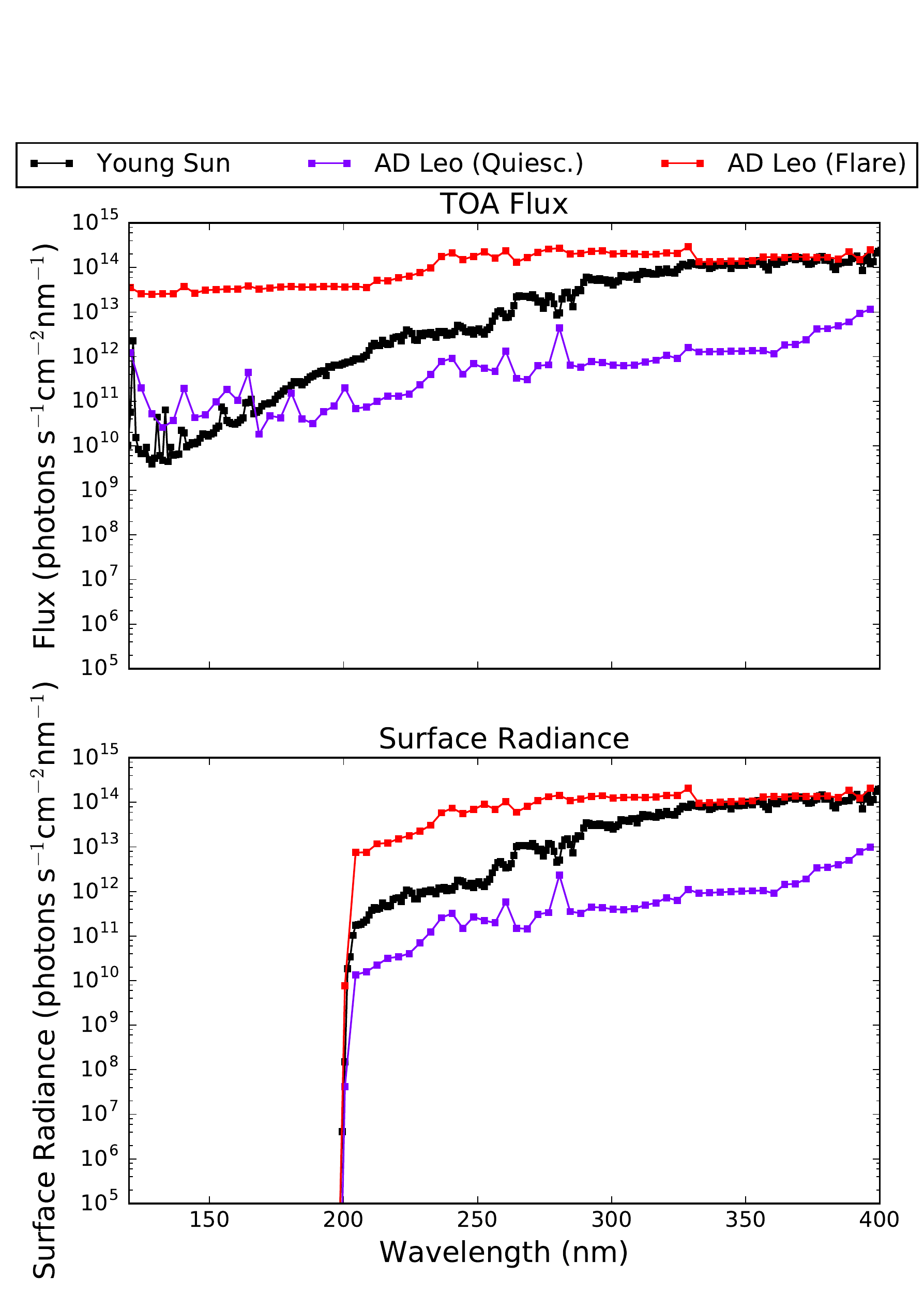}
\caption{TOA fluxes (top) and corresponding surface radiances as a function of wavelength for a prebiotic Earth-analog planet with a 1-bar N$_2$-CO$_2$ atmosphere orbiting AD Leo in quiescence and during the Great Flare of 1985. The young Sun and Earth case (3.9 Ga) is also shown for comparison. \label{fig:results_radiances_flare}}
\end{figure}

Figure~\ref{fig:results_radiances_flare} presents the TOA fluxes for AD Leo in quiescence and at the peak of the Great Flare (912s, \citealt{Segura2010}), and the corresponding surface radiances on a young-Earth-analog planet orbiting it. In flare, AD Leo's UV output rises by orders of magnitude, and the increase in fluence is not confined to line emission as it was for the young Sun in the study of \citet{Cnossen2007}. Rather, the increase is across the UV spectrum, including the continuum. 

\begin{figure}[H]
\centering
\includegraphics[width=10 cm, angle=0]{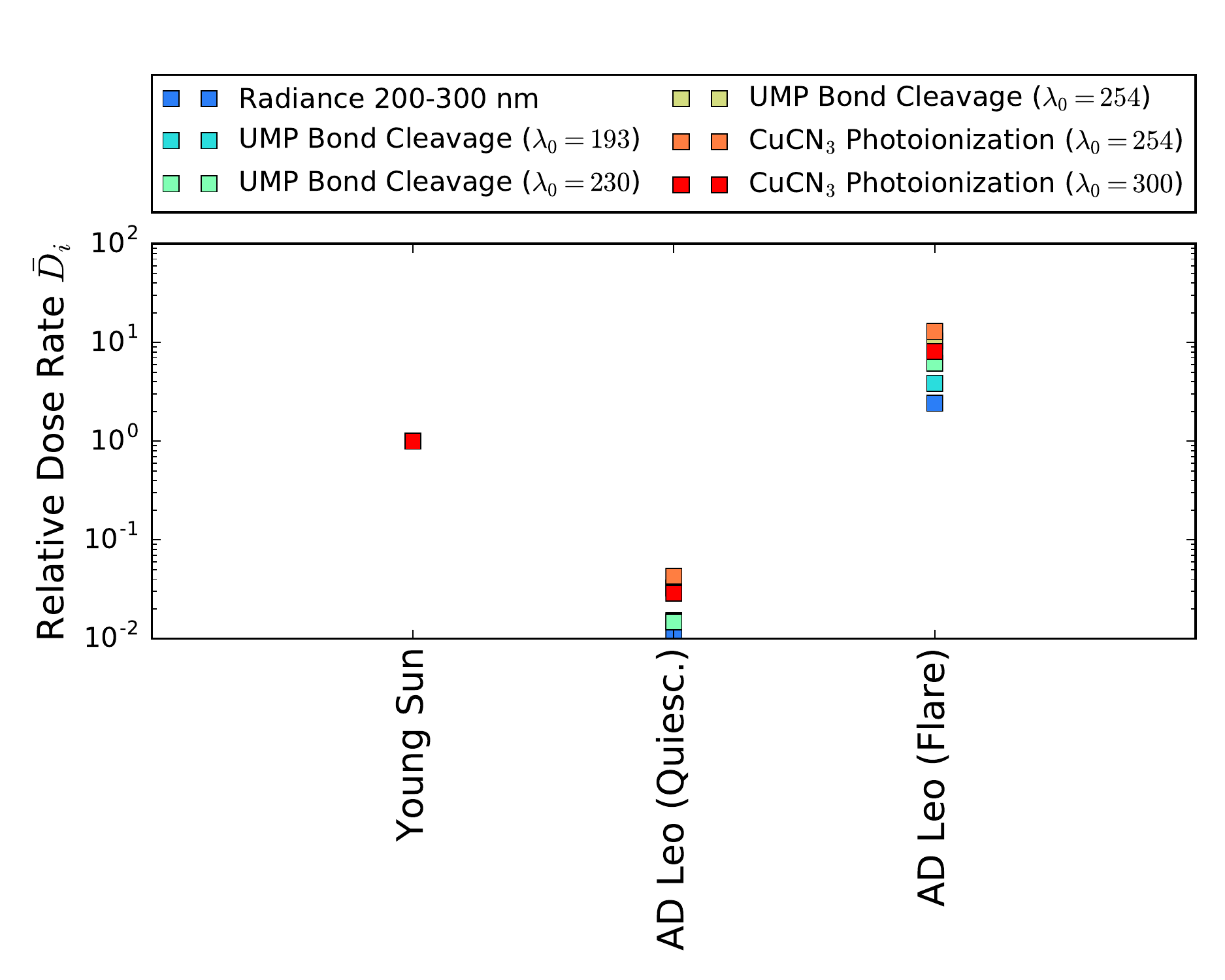}
\caption{UV dose rates $\overline{D}_i$ for UMP-X and CuCN3-Y at the surfaces of a prebiotic Earth-analog planet with a 1-bar N$_2$-CO$_2$ atmosphere orbiting AD Leo in quiescence and during the Great Flare of 1985. Dose rates corresponding to the 3.9 Ga Earth are also shown for comparison. \label{fig:results_doserates_flare}}
\end{figure}

This increase in fluence drives a dramatic increase in reaction rates. Figure~\ref{fig:results_doserates_flare} shows the dose rates for a young-Earth analog orbiting AD Leo in and out of flare, compared to the young Earth. While in quiescence prebiotically relevant reaction rates are suppressed by 1-2 orders of magnitude relative to the young Earth, in flare the reaction rates are enhanced, by up to 1 order of magnitude relative to young Earth. The increase in UV fluence is relatively spectrally flat across the NUV regime that penetrates the atmosphere to the planet surface, and the flare does not deliver radiation that particularly favors or disfavors stressor processes versus eustressor processes relative to quiescence.

\section{Discussion\label{sec:disc}} 
\subsection{Low-UV on M-dwarfs: A Challenge for Abiogenesis?\label{sec:disc_1}}
Prebiotic chemistry on planet orbiting M-dwarfs has access to orders of magnitude less UV radiation than prebiotic chemistry on planets orbiting Solar-type stars, due to lower NUV emission by the cooler M-dwarfs. This poses a challenge for some origin-of-life scenarios on such worlds. As discussed in Section~\ref{sec:intro}, UV light is a key requirement of several proposed prebotic pathways, including the only known pathways for prebiotic syntheses of ribonucleotides  \citep{Powner2009, Ritson2012, Patel2015, Xu2016}. It is possible that these pathways will still function under a low-UV regime, albeit at lower rates. However, it is equally possible that they will not, due to interference or competition with thermal reactions. For example, the core of the sugar synthesis pathway of \citet{Ritson2012} and \citet{Patel2015} involves the reduction of glycolonitrile to glycolaldehyde imine by a photochemically-produced solvated electron, which then hydrolyzes to give the desired glycolaldehyde sugar. However, glycolonitrile can also react with H$_2$S\footnote{Required for the synthesis scenario outlined in \citet{Patel2015}} to produce its alpha-hydroxy thioamide form, which subsequently hydrolyzes to its alpha-hydroxy acid form. If the UV irradiance is low enough, the latter pathway will dominate over the former, and the sugar synthesis pathway will not proceed\footnote{J. Szostak, private communication, 2016 November 6}. Similarly, the efficacy of UV light at selecting and amplifying the population of biogenic ribonucleotides and generating uridine from cytidine in \citet{Powner2009} will be functions of the irradiation level. These examples illustrates the need to characterize the sensitivity of putative prebiotic pathways to the amplitude of irradiation, to understand whether they can proceed under the lower-UV conditions on planets orbiting M-dwarfs. In sum, it is unclear whether UV-dependent prebiotic pathways that may have been important to the origin of life on Earth can proceed in the low-UV environment on planets orbiting M-dwarfs, such as the recently-discovered habitable zone planets Proxima Centauri b \citep{Anglada-Escude2016}, LHS 1140b \citep{Dittmann2017}, or the habitable-zone planets orbiting TRAPPIST-1 \citep{Gillon2017}. 

Even if these pathways do proceed at lower irradiance levels, their reaction rates will likely be orders of magnitude lower than on planets around Sunlike stars. If UV-dependent pathways like the RNA monomer synthesis pathways of \citet{Patel2015} were a rate-limiting step in the origin of life on Earth, then lower M-dwarf NUV radiation could delay the origin of life on planets orbiting M-stars by orders of magnitude. If abiogenesis is fast, this is not a problem. However, if abiogenesis is slow, this could pose a challenge. For example, if abiogenesis took 100 Ma on Earth and UV photochemistry was the rate-limiting step, then it would take $\gtrsim10$ Ga around M-dwarfs, meaning that only old M-dwarfs could host planets with native life. In the worst-case scenario, the Universe might be too young for M-dwarf-orbiting life to have evolved at all.

The UV-scarcity problem may be exacerbated by atmospheric photochemistry, depending on planet history and surface conditions. \citet{Luger2015} show that terrestrial planets with significant initial water inventories orbiting in the current habitable zones of main-sequence M-dwarfs may have undergone a runaway greenhouse and suffered massive loss of water due to enhanced M-dwarf luminosity in their extended pre-main sequence phase, which could have generated high levels of UV-shielding O$_2$ and O$_3$ if not balanced by loss to the mantle and crust. Similarly, \citet{Harman2015} show that Earthlike planets with significant CO$_2$ inventories\footnote{Expected for Earthlike volcanic outgassing} orbiting M-dwarfs can build up abiotic ozone layers comparable in strength to modern Earth's if sinks of O$_2$ and CO are small, because the greater FUV/NUV ratio of M-dwarfs means that O$_3$ production rates are higher and dissociation rates are lower. \citet{Gao2015} demonstrate that CO$_2$-rich, H$_2$O-poor planets orbiting M-dwarfs can also build up abiotic ozone layers comparable to modern Earth's, and for high H levels can also build up H$_2$O$_2$ which can also act as a UV shield at elevated levels. \citet{Wordsworth2014} showed that water-rich planets with low inventories of non-condensible gases like N$_2$ may also abiotically oxygenate.  In sum, planets orbiting M-dwarfs with inefficient sinks of O$_2$ and CO, with desiccated CO$_2$ rich atmospheres, or lacking non-condensible background gases, may photochemically produce atmospheric UV shields like O$_3$ that can further block the already low M-dwarf UV from the planet surface. 

One might argue that our focus on UV light is unwarranted because it is not certain that UV light played a role in the origin of life on Earth, let alone on other planets; indeed, UV light would be altogether absent in the deep-sea hydrothermal vent hypothesis for the origin of life (e.g., \citealt{Martin2008}). Our rationale for focusing on UV light is threefold. First, the nucleotides of RNA show evidence of selection pressure from UV irradiation, suggesting they arose in a UV-rich environment \citep{Rios2013, Beckstead2016, Pollum2016}. If one assumes RNA were the primordial autocatalytic replicator, as is the paradigm in the RNA world hypothesis \citep{Gilbert1986}, this implies UV light was abundant during abiogenesis on Earth. Second, only UV light has been empirically demonstrated to drive the only known prebiotically plausible pathways for the origin of the pyrimidine RNA monomers and for the selective sugar synthesis pathway that sidesteps the asphaltization problem \citep{Patel2015, Benner2012}. It is possible that a different energy source could substitute for UV, but this has not yet been demonstrated in the laboratory. Third, It is difficult to envision sources of free energy that can substitute for UV light in general. The energetic nature of UV light means that it is capable of directly affecting or altering molecular electronic structure, permitting it to effect irreversible changes in the entropic states of molecular systems. To effect similar changes thermally requires sources with temperatures on the orders of thousands of degrees Kelvin, coupled to quenching mechanisms to prevent thermal relaxation to the equilibrium state \citep{Pascal2012}. Two such high-temperature planetary sources are lightning and impacts, which have been contemplated as energy sources for prebiotic molecular synthesis \citep{Chyba1992, Nava-Sedeno2016}. However, these sources face the problem of abundance: lightning and impact-driven shockwaves are estimated to have delivered three orders of magnitude less energy than UV photons ($<300$ nm) to the young Earth \citep{Ferris1975}. Hence, it is difficult to avoid the conclusion that a paucity of UV light might pose a problem for abiogenesis scenarios on M-dwarf planets like Proxima Centauri b, LHS 1140b, or the TRAPPIST-1 planets. Conversely, if life is found on an M-dwarf planet it might imply a pathway to the origin of life very different from what seems to have played out on Earth.

\subsection{Possible Mechanisms to Compensate for Low M-dwarf UV}
We considered whether the potential UV paucity problem could be solved with thinner atmospheres that block less UV light. We might imagine that in some regions of parameter space, elevated M-dwarf EUV emission or interactions with a more intense M-dwarf stellar wind could at least transiently strip the atmosphere from habitable-zone planets, similar to the scenario \citet{Tian2009a} calculated for the young Mars due to elevated early solar EUV emission, or the scenario of \citet{Dong2017b} for the planets of the TRAPPIST-1 system due to enhanced stellar wind interactions. We repeated our study for a tenuous, 1-microbar N$_2$-CO$_2$ atmosphere which provided essentially no attenuation of incoming fluence. Relative dose rates did rise by 1-2 orders of magnitude. However, the strongest increases in reaction rates were for destructive reactions (i.e. UMP photolysis), which is unsurprising: the new fluence admitted by the thinner atmosphere were the destructive FUV ($<200$ nm) wavelengths, making the environment less clement for abiogenesis. In addition, such low atmospheric pressures pose other problems for prebiotic chemistry: for example, liquid water is not stable at pressures below the triple point, i.e. $P_0<6\times10^{-3}$ bar \citep{CRC90}, meaning aqueous phase prebiotic chemistry would face challenges on an airless world. We conclude that atmospheric stripping cannot solve the UV paucity problem for prebiotic chemistry.

Our analysis has focused on main-sequence M-dwarfs. Young M-dwarfs, especially pre-main-sequence M-dwarfs emit a larger fraction of their bolometric luminosity as NUV radiation compared to their main sequence phase \citep{Shkolnik2014}. The lowest-mass M-dwarfs remain in this state for up to $\sim1$ Gyr timescales, and one might speculate whether planets orbiting such stars might receive NUV irradiation more comparable to planets orbiting Sunlike stars. However, the bolometric luminosity of such stars is also higher; planets in the main-sequence habitable zones of such stars are liable to be in a runaway greenhouse state during their pre-main-sequence evolution, with temperatures globally above the boiling point of water and possibly as high as $\gtrsim1000$K \citep{RamirezKaltenegger2014, Luger2015cores, Luger2015, Schaefer2016}. Hence, these objects will lack clement conditions for aqueous-phase prebiotic chemistry, and are unlikely to be habitable during this phase. Planets orbiting farther out, in the pre-main-sequence habitable zone, may remain habitable for up to gigayear timescales \citep{RamirezKaltenegger2014}, and will on average experience NUV irradiation (175-275 nm) up to 1 order of magnitude (OOM) higher than planets orbiting at equivalent bolometric instellations during the star's main sequence phase \citep{Shkolnik2014}. However, these objects will exit the habitable zone when their host star joins the main sequence and reduce their bolometric luminosity. Unless global temperatures are subsequently elevated by mechanisms not considered in the traditional Earth-analog habitable zone calculation, e.g. greenhouse warming due to elevated levels of atmospheric H$_2$ \citep{Stevenson1999, Pierrehumbert2011, Wordsworth2013h2, Wordsworth2017, RamirezKaltenegger2017}, liquid water will freeze at the planet surface and life will be confined to volcanic or subsurface reservoirs, whose ability to produce global atmospheric biosignatures is unclear. More critically, it is uncertain whether a 1-OOM enhancement\footnote{We note that the enhancement in NUV luminosity relative to bolometric luminosity found by \citet{Shkolnik2014} for young M-dwarfs is 1 OOM \textit{on average}. It is possible that some subset of M-dwarfs will show even higher enhancements; if so, UV-dependent prebiotic chemistry on planets orbiting such objects will face a proportionately smaller UV paucity problem, though the other challenges faced by prebiotic chemistry on such objects will remain.}  in 175-275 nm fluence is adequate to compensate for the 2-4 OOM suppression in bioactive NUV fluence on M-dwarf planets compared to the early Earth; experimental characterization of the fluence dependence of UV-sensitive putative prebiotic chemistry remains necessary to answer this question. Nevertheless, planets orbiting in the habitable zones of late-type (low-mass) pre-main-sequence M-dwarfs remain of special interest from the perspective of UV-sensitive prebiotic chemistry.

We considered whether M-dwarf flares, which are typically considered barriers to habitability \citep{Segura2010}, might provide a partial solution to the problem of UV paucity on M-dwarf planets \citep{Buccino2007}. During the Great Flare of 1985, AD Leo's NUV output increased dramatically, delivering more NUV radiation to the surface of a young-Earth-analog planet than the young Sun for a period of at least 0.7 hour and possibly longer\footnote{The flare lasted 4 hours but we located NUV observations for only the first 0.7 hours; see \citealt{Segura2010}}. Flare frequency studies of AD Leo suggest that flares of this strength occur with frequency $0.1$ day$^{-1}$ \citet{Pettersen1984} \footnote{Lower-energy flares occur more frequently but emit substantially less UV radiation, and are likely also shorter in duration (c.f. the study of \citealt{Hawley2014} for GJ 1243)}. For comparison, the typical irradiation time used in laboratory studies of UV-sensitive prebiotic pathways like C-to-U conversion and glycolaldehyde/glyceraldehyde formation at fluence levels comparable with the young Sun is on the order of 4 hours \citep{Todd2017a}, a duration comparable to strong M-dwarf flares. One might imagine a scenario whereby photosensitive prebiotic chemistry proceeds during the high-UV flares and ceases during stellar quiescence, providing an activity-powered analog to the terrestrial day/night cycle that would be absent on tidally-locked habitable zone M-dwarf planets. Experimental studies are required to constrain whether shorter duration but higher intensity irradiation corresponding to an M-dwarf flare is sufficient to power putative prebiotic chemistry. Theoretical studies are also required to understand whether the enhanced UV radiation these flares provide also strip the atmosphere, which would negatively impact planetary habitability and obviate this solution. We note that this mechanism could best solve the UV-paucity problem for planets orbiting the most active M-stars, like AD Leo. Planets orbiting active stars would experience much lower flare rates; for example, extrapolation of white-light flare frequency studies for Proxima Centauri suggest that Proxima Centauri b would experience UV instellation comparable the Great Flare of AD Leo only $\sim8$ year$^{-1}$, $5\times$ less frequently than an Earth-analog planet orbiting AD Leo \citep{Davenport2016}.

\subsection{Laboratory Follow-Up}
Our work suggests a critical need for laboratory studies to determine whether putative UV-dependent prebiotic pathways (e.g., \citealt{Powner2009}, \citealt{Ritson2012}, \citealt{Patel2015}, \citealt{Xu2016}) could function at the $10-1000\times$ lower UV irradiation accessible on planets orbiting M-dwarfs compared to the young Earth. First, the fluence dependence of these pathways should be determined: How do their rates vary as a function of irradiation level, and at which minimum fluence do they shut down? These questions should be probed down to fluence levels as low as $10^{-3}-10^{-4}$ of that accessible on early Earth, corresponding to the fluence available on quiet M-dwarfs. Integrated from 200-300 nm, prebiotic chemistry on early Earth would have been exposed to $\sim6\times10^{14}$ photons s$^{-1}$cm$^{-2}$, implying the need to probe down to photon fluxes as low as $\sim6\times10^{11}$ photons s$^{-1}$cm$^{-2}$ ($5$ erg s$^{-1}$cm$^{-2}$ at the 254 nm line generally used in prebiotic chemistry studies). 

If such studies determine that one or more of these pathways fail at M-dwarf irradiation levels, experiments should be done to determine whether transient high UV irradiation due to flares can substitute for steady-state irradiation. The best-established case to simulate is the Great Flare on AD Leo; this event delivered $2\times10^{15}$ photons/s/cm$^2$ ($2\times10^{4}$ erg s$^{-1}$cm$^{-2}$ at 254 nm) to the surface of an orbiting young-Earth analog, integrated from 200-300 nm. Such fluence levels would have been available for periods of 0.7-4 hours at intervals of $\approx240$ hours according to the power-law presented in \citet{Pettersen1984}. If such intense flares are too infrequent to power prebiotic chemistry, then the influence of lower-intensity, higher-frequency flares may be considered. For example, flares 1/10 as energetic as the Great Flare occur every $60$ hours (see \ref{sec:appendix_ffd}). If flare frequencies $\sim5\times$ less than that of AD Leo are found to be sufficient to power UV-dependent prebiotic chemistry, then flare instellation could solve the potential UV paucity problem for Proxima Centauri b.

\section{Conclusions\label{sec:conc}}
Recent laboratory studies suggest that UV light played a critical role in the origin of life on Earth. We have used a radiative transfer model to evaluate the UV surface environment on planets analogous to prebiotic Earth (N$_2$-CO$_2$ atmosphere) orbiting M-dwarfs, and used action spectra to calculate dose rates and quantify the implications for prebiotic chemistry on such worlds. Such planets are the most compelling target for biosignature searches over at least the next decade. 

We find the UV surface environment on M-dwarf planets is chiefly differentiated from planets orbiting Sunlike stars in that they have access to orders of magnitude less prebiotically-useful NUV radiation, due to the lower emission of M-dwarfs at these wavelengths. Planets orbiting in the transient habitable zones of pre-main-sequence late-type M-dwarfs should experience more NUV irradiation, but only transiently, and not enough to close the deficit with planets orbiting Sunlike stars. This raises uncertainty over whether the UV-dependent prebiotic pathways that may have lead to the origin of life on Earth could function on planets orbiting M-dwarfs, such as the recently-discovered habitable-zone planets orbiting Proxima Centauri, LHS 1140, and TRAPPIST-1. Even if the pathways proceed, their reaction rates will likely be orders of magnitude lower than for planets around Sunlike stars, potentially slowing abiogenesis. 

These scenarios can be tested empirically, through laboratory studies to measure  the reaction rate of putative UV-dependent prebiotic pathways  (e.g. \citealt{Ritson2012}, \citealt{Patel2015}, \citealt{Xu2016}) to the amplitude of irradiation. Such laboratory studies are urgently needed, in order to identify the most compelling targets for biosignature search with next-generation near-term instruments like JWST and the ELTs. If such laboratory studies reveal that near-solar levels of NUV fluence are required to move forward UV-sensitive prebiotic chemistry, they will raise questions regarding the prospects for the emergence of life on M-dwarf planets. Interestingly, in this case planets orbiting active M-dwarfs may be more compelling candidates for abiogenesis scenarios, due to both the higher quiescent emission of such stars and the frequent flares from such stars, which will periodically illuminate the planet with elevated levels of UV that may power prebiotic photochemistry. Laboratory experiments are required to understand whether burst of brief ($\sim$hours), high-intensity radiation separated by long intervals ($\sim10$ days) can substitute for steady-state solar emission. Theoretical studies are required to ensure that enhanced UV and particle fluxes from such flares would not also strip the planetary atmosphere, obviating this solution. 

\section{Acknowledgements}
We thank Z. Todd, J. Szostak, and A. Beckstead for their insight regarding prebiotic chemistry and its interaction with UV light. We thank S. Rugheimer, A. Segura, K. France, S. Engle, C. Johns-Krull, R. Osten, and C. Bonfio for sharing their data with us, and for their answers to our questions. We thank L. Walkowicz, V. Meadows, O. Venot, and R. Loyd for insightful discussions. This research has made use of NASAÕs Astrophysics Data System Bibliographic Services, and the MPI-Mainz UV-VIS Spectral Atlas of Gaseous Molecules.

S.R. and D.D.S. gratefully acknowledge support from the Simons Foundation, grant no. 290360. R. W. acknowledges funding from NASA HW Grant NNX16AR86G.

\clearpage

\appendix

\section{Flare Frequency Distribution Calculation \label{sec:appendix_ffd}}

In this section, we calculate explicitly the flare frequencies alluded to in Section~\ref{sec:disc}

For AD Leo: \citet{Pettersen1984} find the flare frequency distribution to be fit by $$\log[\nu(E_{U_0})]=\alpha-\beta\log[E_{U_{0}}], $$ where $\nu$ is the frequency of flares with integrated U-band energy $E_U\geq E_{U_0}$ in units of s$^{-1}$, $\alpha=15.0 \pm 2.1$, and $\beta=0.62\pm0.09$. The AD Leo Great Flare of 1985 had a U-band energy of $10^{33.8}$ erg (\citealt{Hawley1991}, Table 8), corresponding to a frequency of $\nu=10^{\alpha-\beta\times33.8} s^{-1}=1.1\times10^{-6} s^{-1}=0.1 day^{-1}$.  Lower energy flares occur more frequently: A flare with $1/10$ the Great Flare's U-band energy would occur with frequency $\nu=10^{\alpha-\beta\times32.8} s^{-1}=4.6\times10^{-6} s^{-1}=0.4$ day$^{-1}$.

\clearpage

\bibliography{mdwarf_v5.bib}{}

\begin{thebibliography}{}
\expandafter\ifx\csname natexlab\endcsname\relax\def\natexlab#1{#1}\fi
\providecommand{\url}[1]{\href{#1}{#1}}

\bibitem[{{Anglada-Escud{\'e}} {et~al.}(2016){Anglada-Escud{\'e}}, {Amado},
  {Barnes}, {Berdi{\~n}as}, {Butler}, {Coleman}, {de La Cueva}, {Dreizler},
  {Endl}, {Giesers}, {Jeffers}, {Jenkins}, {Jones}, {Kiraga}, {K{\"u}rster},
  {L{\'o}pez-Gonz{\'a}lez}, {Marvin}, {Morales}, {Morin}, {Nelson}, {Ortiz},
  {Ofir}, {Paardekooper}, {Reiners}, {Rodr{\'{\i}}guez},
  {Rodr{\'{\i}}guez-L{\'o}pez}, {Sarmiento}, {Strachan}, {Tsapras}, {Tuomi}, \&
  {Zechmeister}}]{Anglada-Escude2016}
{Anglada-Escud{\'e}}, G., {Amado}, P.~J., {Barnes}, J., {et~al.} 2016, \nat,
  536, 437

\bibitem[{{Batalha} {et~al.}(2015){Batalha}, {Domagal-Goldman}, {Ramirez}, \&
  {Kasting}}]{Batalha2015}
{Batalha}, N., {Domagal-Goldman}, S.~D., {Ramirez}, R., \& {Kasting}, J.~F.
  2015, \icarus, 258, 337

\bibitem[{Beckstead {et~al.}(2016)Beckstead, Zhang, de~Vries, \&
  Kohler}]{Beckstead2016}
Beckstead, A.~A., Zhang, Y., de~Vries, M.~S., \& Kohler, B. 2016, Physical
  Chemistry Chemical Physics, 18, 24228

\bibitem[{Benner {et~al.}(2012)Benner, Kim, \& Carrigan}]{Benner2012}
Benner, S.~A., Kim, H.-J., \& Carrigan, M.~A. 2012, Accounts of chemical
  research, 45, 2025

\bibitem[{{Bourrier} {et~al.}(2017){Bourrier}, {Ehrenreich}, {Wheatley},
  {Bolmont}, {Gillon}, {de Wit}, {Burgasser}, {Jehin}, {Queloz}, \&
  {Triaud}}]{Bourrier2017}
{Bourrier}, V., {Ehrenreich}, D., {Wheatley}, P.~J., {et~al.} 2017, \aap, 599,
  L3

\bibitem[{{Buccino} {et~al.}(2007){Buccino}, {Lemarchand}, \&
  {Mauas}}]{Buccino2007}
{Buccino}, A.~P., {Lemarchand}, G.~A., \& {Mauas}, P.~J.~D. 2007, \icarus, 192,
  582

\bibitem[{{Catling} {et~al.}(2017){Catling}, {Krissansen-Totton}, {Kiang},
  {Crisp}, {Robinson}, {DasSarma}, {Rushby}, {Del Genio}, {Bains}, \&
  {Domagal-Goldman}}]{Catling2017}
{Catling}, D.~C., {Krissansen-Totton}, J., {Kiang}, N.~Y., {et~al.} 2017, ArXiv
  e-prints, arXiv:1705.06381

\bibitem[{{Chyba} \& {Sagan}(1992)}]{Chyba1992}
{Chyba}, C., \& {Sagan}, C. 1992, \nat, 355, 125

\bibitem[{{Claire} {et~al.}(2012){Claire}, {Sheets}, {Cohen}, {Ribas},
  {Meadows}, \& {Catling}}]{Claire2012}
{Claire}, M.~W., {Sheets}, J., {Cohen}, M., {et~al.} 2012, \apj, 757, 95

\bibitem[{{Cnossen} {et~al.}(2007){Cnossen}, {Sanz-Forcada}, {Favata},
  {Witasse}, {Zegers}, \& {Arnold}}]{Cnossen2007}
{Cnossen}, I., {Sanz-Forcada}, J., {Favata}, F., {et~al.} 2007, Journal of
  Geophysical Research (Planets), 112, 2008

\bibitem[{Cockell(1999)}]{Cockell1999}
Cockell, C.~S. 1999, Icarus, 141, 399

\bibitem[{{Cockell}(2000{\natexlab{a}})}]{Cockell2000UVhist}
{Cockell}, C.~S. 2000{\natexlab{a}}, \planss, 48, 203

\bibitem[{{Cockell}(2000{\natexlab{b}})}]{Cockell2000oceans}
---. 2000{\natexlab{b}}, Origins of Life and Evolution of the Biosphere, 30,
  467

\bibitem[{Cockell(2002)}]{Cockell2002}
Cockell, C.~S. 2002, International Journal of Astrobiology, 1, 31

\bibitem[{{Cowan} {et~al.}(2015){Cowan}, {Greene}, {Angerhausen}, {Batalha},
  {Clampin}, {Col{\'o}n}, {Crossfield}, {Fortney}, {Gaudi}, {Harrington},
  {Iro}, {Lillie}, {Linsky}, {Lopez-Morales}, {Mandell}, \&
  {Stevenson}}]{Cowan2015}
{Cowan}, N.~B., {Greene}, T., {Angerhausen}, D., {et~al.} 2015, \pasp, 127, 311

\bibitem[{Cronin(2014)}]{Cronin2014}
Cronin, T.~W. 2014, Journal of the Atmospheric Sciences, 71, 2994

\bibitem[{{Davenport} {et~al.}(2016){Davenport}, {Kipping}, {Sasselov},
  {Matthews}, \& {Cameron}}]{Davenport2016}
{Davenport}, J.~R.~A., {Kipping}, D.~M., {Sasselov}, D., {Matthews}, J.~M., \&
  {Cameron}, C. 2016, \apjl, 829, L31

\bibitem[{{Dittmann} {et~al.}(2017){Dittmann}, {Irwin}, {Charbonneau},
  {Bonfils}, {Astudillo-Defru}, {Haywood}, {Berta-Thompson}, {Newton},
  {Rodriguez}, {Winters}, {Tan}, {Almenara}, {Bouchy}, {Delfosse}, {Forveille},
  {Lovis}, {Murgas}, {Pepe}, {Santos}, {Udry}, {W{\"u}nsche}, {Esquerdo},
  {Latham}, \& {Dressing}}]{Dittmann2017}
{Dittmann}, J.~A., {Irwin}, J.~M., {Charbonneau}, D., {et~al.} 2017, \nat, 544,
  333

\bibitem[{{Dong} {et~al.}(2017){Dong}, {Jin}, {Lingam}, {Airapetian}, {Ma}, \&
  {van der Holst}}]{Dong2017b}
{Dong}, C., {Jin}, M., {Lingam}, M., {et~al.} 2017, ArXiv e-prints,
  arXiv:1705.05535

\bibitem[{{Dressing} \& {Charbonneau}(2015)}]{Dressing2015}
{Dressing}, C.~D., \& {Charbonneau}, D. 2015, \apj, 807, 45

\bibitem[{Ferris \& Chen(1975)}]{Ferris1975}
Ferris, J., \& Chen, C. 1975, Journal of the American Chemical Society, 97,
  2962

\bibitem[{{France} {et~al.}(2016){France}, {Parke Loyd}, {Youngblood}, {Brown},
  {Schneider}, {Hawley}, {Froning}, {Linsky}, {Roberge}, {Buccino},
  {Davenport}, {Fontenla}, {Kaltenegger}, {Kowalski}, {Mauas}, {Miguel},
  {Redfield}, {Rugheimer}, {Tian}, {Vieytes}, {Walkowicz}, \&
  {Weisenburger}}]{France2016}
{France}, K., {Parke Loyd}, R.~O., {Youngblood}, A., {et~al.} 2016, \apj, 820,
  89

\bibitem[{{Gao} {et~al.}(2015){Gao}, {Hu}, {Robinson}, {Li}, \&
  {Yung}}]{Gao2015}
{Gao}, P., {Hu}, R., {Robinson}, T.~D., {Li}, C., \& {Yung}, Y.~L. 2015, \apj,
  806, 249

\bibitem[{Gilbert(1986)}]{Gilbert1986}
Gilbert, W. 1986, Nature, 319, 618.
\newblock \url{http://www.citeulike.org/group/244/article/400674}

\bibitem[{{Gillon} {et~al.}(2017){Gillon}, {Triaud}, {Demory}, {Jehin}, {Agol},
  {Deck}, {Lederer}, {de Wit}, {Burdanov}, {Ingalls}, {Bolmont}, {Leconte},
  {Raymond}, {Selsis}, {Turbet}, {Barkaoui}, {Burgasser}, {Burleigh}, {Carey},
  {Chaushev}, {Copperwheat}, {Delrez}, {Fernandes}, {Holdsworth}, {Kotze}, {Van
  Grootel}, {Almleaky}, {Benkhaldoun}, {Magain}, \& {Queloz}}]{Gillon2017}
{Gillon}, M., {Triaud}, A.~H.~M.~J., {Demory}, B.-O., {et~al.} 2017, \nat, 542,
  456

\bibitem[{Gurzadyan \& G{\"o}rner(1994)}]{Gurzadyan1994}
Gurzadyan, G.~G., \& G{\"o}rner, H. 1994, Photochemistry and photobiology, 60,
  323

\bibitem[{{Harman} {et~al.}(2015){Harman}, {Schwieterman}, {Schottelkotte}, \&
  {Kasting}}]{Harman2015}
{Harman}, C.~E., {Schwieterman}, E.~W., {Schottelkotte}, J.~C., \& {Kasting},
  J.~F. 2015, \apj, 812, 137

\bibitem[{{Hawley} {et~al.}(2014){Hawley}, {Davenport}, {Kowalski},
  {Wisniewski}, {Hebb}, {Deitrick}, \& {Hilton}}]{Hawley2014}
{Hawley}, S.~L., {Davenport}, J.~R.~A., {Kowalski}, A.~F., {et~al.} 2014, \apj,
  797, 121

\bibitem[{{Hawley} \& {Pettersen}(1991)}]{Hawley1991}
{Hawley}, S.~L., \& {Pettersen}, B.~R. 1991, \apj, 378, 725

\bibitem[{{Heath} {et~al.}(1999){Heath}, {Doyle}, {Joshi}, \&
  {Haberle}}]{Heath1999}
{Heath}, M.~J., {Doyle}, L.~R., {Joshi}, M.~M., \& {Haberle}, R.~M. 1999,
  Origins of Life and Evolution of the Biosphere, 29, 405

\bibitem[{{Kasting}(1991)}]{Kasting1991}
{Kasting}, J.~F. 1991, \icarus, 94, 1

\bibitem[{{Kasting} {et~al.}(1993){Kasting}, {Whitmire}, \&
  {Reynolds}}]{Kasting1993habzone}
{Kasting}, J.~F., {Whitmire}, D.~P., \& {Reynolds}, R.~T. 1993, \icarus, 101,
  108

\bibitem[{{Kiang} {et~al.}(2007){Kiang}, {Segura}, {Tinetti}, {Govindjee},
  {Blankenship}, {Cohen}, {Siefert}, {Crisp}, \& {Meadows}}]{Kiang2007}
{Kiang}, N.~Y., {Segura}, A., {Tinetti}, G., {et~al.} 2007, Astrobiology, 7,
  252

\bibitem[{{Kowalski} {et~al.}(2013){Kowalski}, {Hawley}, {Wisniewski}, {Osten},
  {Hilton}, {Holtzman}, {Schmidt}, \& {Davenport}}]{Kowalski2013}
{Kowalski}, A.~F., {Hawley}, S.~L., {Wisniewski}, J.~P., {et~al.} 2013, \apjs,
  207, 15

\bibitem[{{Lammer} {et~al.}(2009){Lammer}, {Bredeh{\"o}ft}, {Coustenis},
  {Khodachenko}, {Kaltenegger}, {Grasset}, {Prieur}, {Raulin}, {Ehrenfreund},
  {Yamauchi}, {Wahlund}, {Grie{\ss}meier}, {Stangl}, {Cockell}, {Kulikov},
  {Grenfell}, \& {Rauer}}]{Lammer2009}
{Lammer}, H., {Bredeh{\"o}ft}, J.~H., {Coustenis}, A., {et~al.} 2009, \aapr,
  17, 181

\bibitem[{Lide(2009)}]{CRC90}
Lide, D.~R., ed. 2009, CRC Handbook of Chemistry and Physics, 90th edn. (Boca
  Raton, FL: CRC Press)

\bibitem[{{Loyd} {et~al.}(2016){Loyd}, {France}, {Youngblood}, {Schneider},
  {Brown}, {Hu}, {Linsky}, {Froning}, {Redfield}, {Rugheimer}, \&
  {Tian}}]{Loyd2016}
{Loyd}, R.~O.~P., {France}, K., {Youngblood}, A., {et~al.} 2016, \apj, 824, 102

\bibitem[{{Luger} \& {Barnes}(2015)}]{Luger2015}
{Luger}, R., \& {Barnes}, R. 2015, Astrobiology, 15, 119

\bibitem[{{Luger} {et~al.}(2015){Luger}, {Barnes}, {Lopez}, {Fortney},
  {Jackson}, \& {Meadows}}]{Luger2015cores}
{Luger}, R., {Barnes}, R., {Lopez}, E., {et~al.} 2015, Astrobiology, 15, 57

\bibitem[{Magnani(2015)}]{Magnani2015}
Magnani, C.~J. 2015, {A.B. Thesis}, Harvard University

\bibitem[{Martin {et~al.}(2008)Martin, Baross, Kelley, \& Russell}]{Martin2008}
Martin, W., Baross, J., Kelley, D., \& Russell, M.~J. 2008, Nature Reviews
  Microbiology, 6, 805

\bibitem[{{Meadows} {et~al.}(2016){Meadows}, {Arney}, {Schwieterman},
  {Lustig-Yaeger}, {Lincowski}, {Robinson}, {Domagal-Goldman}, {Barnes},
  {Fleming}, {Deitrick}, {Luger}, {Driscoll}, {Quinn}, \&
  {Crisp}}]{Meadows2016}
{Meadows}, V.~S., {Arney}, G.~N., {Schwieterman}, E.~W., {et~al.} 2016, ArXiv
  e-prints, arXiv:1608.08620

\bibitem[{Mulkidjanian {et~al.}(2003)Mulkidjanian, Cherepanov, \&
  Galperin}]{Mulkidjanian2003}
Mulkidjanian, A.~Y., Cherepanov, D.~A., \& Galperin, M.~Y. 2003, BMC
  Evolutionary Biology, 3, 12.
\newblock \url{http://www.biomedcentral.com/1471-2148/3/12/}

\bibitem[{{Nava-Sede{\~n}o} {et~al.}(2016){Nava-Sede{\~n}o}, {Ortiz-Cervantes},
  {Segura}, \& {Domagal-Goldman}}]{Nava-Sedeno2016}
{Nava-Sede{\~n}o}, J.~M., {Ortiz-Cervantes}, A., {Segura}, A., \&
  {Domagal-Goldman}, S.~D. 2016, Astrobiology, 16, 744

\bibitem[{{O'Malley-James} \& {Kaltenegger}(2016)}]{Omalley-James2016}
{O'Malley-James}, J.~T., \& {Kaltenegger}, L. 2016, ArXiv e-prints,
  arXiv:1608.06930

\bibitem[{{Osten}(2016)}]{Osten2016}
{Osten}, R. 2016, Heliophysics: Active Stars, their Astrospheres, and Impacts
  on Planetary Environments, ed. C.~J. Schrijver, F.~Bagenal, \& J.~J. Sojka
  (Cambridge University Press)

\bibitem[{Pascal(2012)}]{Pascal2012}
Pascal, R. 2012, Journal of Systems Chemistry, 3, 1

\bibitem[{Patel {et~al.}(2015)Patel, Percivalle, Ritson, Duffy, \&
  Sutherland}]{Patel2015}
Patel, B.~H., Percivalle, C., Ritson, D.~J., Duffy, C.~D., \& Sutherland, J.~D.
  2015, Nature Chemistry, 1.
\newblock \url{http://www.nature.com/doifinder/10.1038/nchem.2202}

\bibitem[{{Pettersen} {et~al.}(1984){Pettersen}, {Coleman}, \&
  {Evans}}]{Pettersen1984}
{Pettersen}, B.~R., {Coleman}, L.~A., \& {Evans}, D.~S. 1984, \apjs, 54, 375

\bibitem[{{Pierrehumbert} \& {Gaidos}(2011)}]{Pierrehumbert2011}
{Pierrehumbert}, R., \& {Gaidos}, E. 2011, \apjl, 734, L13

\bibitem[{Pollum {et~al.}(2016)Pollum, Ashwood, Jockusch, Lam, \&
  Crespo-Hern{\'a}ndez}]{Pollum2016}
Pollum, M., Ashwood, B., Jockusch, S., Lam, M., \& Crespo-Hern{\'a}ndez, C.~E.
  2016, Journal of the American Chemical Society, 138, 11457

\bibitem[{Powner {et~al.}(2009)Powner, Gerland, \& Sutherland}]{Powner2009}
Powner, M.~W., Gerland, B., \& Sutherland, J.~D. 2009, Nature, 459, 239.
\newblock \url{http://www.ncbi.nlm.nih.gov/pubmed/19444213}

\bibitem[{{Ramirez} \& {Kaltenegger}(2017)}]{RamirezKaltenegger2017}
{Ramirez}, R., \& {Kaltenegger}, L. 2017, ArXiv e-prints, arXiv:1702.08618

\bibitem[{{Ramirez} \& {Kaltenegger}(2014)}]{RamirezKaltenegger2014}
{Ramirez}, R.~M., \& {Kaltenegger}, L. 2014, \apjl, 797, L25

\bibitem[{{Ranjan} \& {Sasselov}(2016)}]{Ranjan2016}
{Ranjan}, S., \& {Sasselov}, D.~D. 2016, Astrobiology, 16, 68

\bibitem[{{Ranjan} \& {Sasselov}(2017)}]{Ranjan2017a}
---. 2017, Astrobiology, 17, 169

\bibitem[{{Ranjan} {et~al.}(2017){Ranjan}, {Wordsworth}, \&
  {Sasselov}}]{Ranjan2017b}
{Ranjan}, S., {Wordsworth}, R.~D., \& {Sasselov}, D.~D. 2017, ArXiv e-prints,
  arXiv:1701.01373

\bibitem[{Ranjan {et~al.}(2017)Ranjan, Wordsworth, \& Sasselov}]{ZenodoRelease}
Ranjan, S., Wordsworth, R.~D., \& Sasselov, D.~D. 2017, {Code to reproduce
  Ranjan et al. (2017), ApJ, submitted}, v1.0.1,  Zenodo, {If making use of
  this code or the generated auxiliary files, please cite the descriptor paper:
  Ranjan, S, Wordsworth, Robin D. and Sasselov, Dimitar D. "The Surface UV
  Environment on Prebiotic Planets Orbiting M-dwarfs: Implications for
  Prebiotic Chemistry \& Need for Experimental Follow-Up". Submitted to the
  Astrophysical Journal (2017). arXiv: 1705.02350}, doi:10.5281/zenodo.584133.
\newblock \url{https://doi.org/10.5281/zenodo.584133}

\bibitem[{Rios \& Tor(2013)}]{Rios2013}
Rios, A.~C., \& Tor, Y. 2013, Israel journal of chemistry, 53, 469

\bibitem[{Ritson \& Sutherland(2012)}]{Ritson2012}
Ritson, D., \& Sutherland, J.~D. 2012, Nature chemistry, 4, 895.
\newblock
  \url{http://www.pubmedcentral.nih.gov/articlerender.fcgi?artid=3589744\&tool=pmcentrez\&rendertype=abstract}

\bibitem[{{Rodler} \& {L{\'o}pez-Morales}(2014)}]{Rodler2014}
{Rodler}, F., \& {L{\'o}pez-Morales}, M. 2014, \apj, 781, 54

\bibitem[{{Rojas-Ayala} {et~al.}(2012){Rojas-Ayala}, {Covey}, {Muirhead}, \&
  {Lloyd}}]{Rojas-Ayala2012}
{Rojas-Ayala}, B., {Covey}, K.~R., {Muirhead}, P.~S., \& {Lloyd}, J.~P. 2012,
  \apj, 748, 93

\bibitem[{Ront{\'{o}} {et~al.}(2003)Ront{\'{o}}, B{\'{e}}rces, Lammer, Cockell,
  Molina-Cuberos, Patel, Selsis, \& Be}]{Ronto2003}
Ront{\'{o}}, G., B{\'{e}}rces, A., Lammer, H., {et~al.} 2003, Photochemistry
  and photobiology, 77, 34.
\newblock \url{http://www.ncbi.nlm.nih.gov/pubmed/12856880}

\bibitem[{Rosenberg {et~al.}(2008)Rosenberg, {Abu Haija}, \&
  Ryan}]{Rosenberg2008}
Rosenberg, R., {Abu Haija}, M., \& Ryan, P. 2008, Physical Review Letters, 101,
  178301.
\newblock \url{http://link.aps.org/doi/10.1103/PhysRevLett.101.178301}

\bibitem[{{Rugheimer} {et~al.}(2015){Rugheimer}, {Segura}, {Kaltenegger}, \&
  {Sasselov}}]{Rugheimer2015}
{Rugheimer}, S., {Segura}, A., {Kaltenegger}, L., \& {Sasselov}, D. 2015, \apj,
  806, 137

\bibitem[{Sagan(1973)}]{Sagan1973}
Sagan, C. 1973, Journal of theoretical biology, 39, 195.
\newblock \url{http://www.ncbi.nlm.nih.gov/pubmed/4741712}

\bibitem[{Sarker {et~al.}(2013)Sarker, Takahashi, Obayashi, Kaneko, \&
  Kobayashi}]{Sarker2013}
Sarker, P.~K., Takahashi, J.-i., Obayashi, Y., Kaneko, T., \& Kobayashi, K.
  2013, Advances in Space Research, 51, 2235.
\newblock \url{http://linkinghub.elsevier.com/retrieve/pii/S0273117713000665}

\bibitem[{{Scalo} {et~al.}(2007){Scalo}, {Kaltenegger}, {Segura}, {Fridlund},
  {Ribas}, {Kulikov}, {Grenfell}, {Rauer}, {Odert}, {Leitzinger}, {Selsis},
  {Khodachenko}, {Eiroa}, {Kasting}, \& {Lammer}}]{Scalo2007}
{Scalo}, J., {Kaltenegger}, L., {Segura}, A.~G., {et~al.} 2007, Astrobiology,
  7, 85

\bibitem[{{Schaefer} {et~al.}(2016){Schaefer}, {Wordsworth}, {Berta-Thompson},
  \& {Sasselov}}]{Schaefer2016}
{Schaefer}, L., {Wordsworth}, R.~D., {Berta-Thompson}, Z., \& {Sasselov}, D.
  2016, \apj, 829, 63

\bibitem[{{Seager}(2014)}]{Seager2014}
{Seager}, S. 2014, Proceedings of the National Academy of Science, 111, 12634

\bibitem[{{Segura} {et~al.}(2005){Segura}, {Kasting}, {Meadows}, {Cohen},
  {Scalo}, {Crisp}, {Butler}, \& {Tinetti}}]{Segura2005}
{Segura}, A., {Kasting}, J.~F., {Meadows}, V., {et~al.} 2005, Astrobiology, 5,
  706

\bibitem[{{Segura} {et~al.}(2003){Segura}, {Krelove}, {Kasting}, {Sommerlatt},
  {Meadows}, {Crisp}, {Cohen}, \& {Mlawer}}]{Segura2003}
{Segura}, A., {Krelove}, K., {Kasting}, J.~F., {et~al.} 2003, Astrobiology, 3,
  689

\bibitem[{{Segura} {et~al.}(2010){Segura}, {Walkowicz}, {Meadows}, {Kasting},
  \& {Hawley}}]{Segura2010}
{Segura}, A., {Walkowicz}, L.~M., {Meadows}, V., {Kasting}, J., \& {Hawley}, S.
  2010, Astrobiology, 10, 751

\bibitem[{Setlow(1974)}]{Setlow1974}
Setlow, R.~B. 1974, Proceedings of the National Academy of Sciences, 71, 3363

\bibitem[{{Shields} {et~al.}(2016){Shields}, {Ballard}, \&
  {Johnson}}]{Shields2016}
{Shields}, A.~L., {Ballard}, S., \& {Johnson}, J.~A. 2016, ArXiv e-prints,
  arXiv:1610.05765

\bibitem[{{Shkolnik} {et~al.}(2009){Shkolnik}, {Liu}, \& {Reid}}]{Shkolnik2009}
{Shkolnik}, E., {Liu}, M.~C., \& {Reid}, I.~N. 2009, \apj, 699, 649

\bibitem[{{Shkolnik} \& {Barman}(2014)}]{Shkolnik2014}
{Shkolnik}, E.~L., \& {Barman}, T.~S. 2014, \aj, 148, 64

\bibitem[{{\v{S}}poner {et~al.}(2016){\v{S}}poner, Szabla, G{\'o}ra, Saitta,
  Pietrucci, Saija, Di~Mauro, Saladino, Ferus, Civi{\v{s}},
  {et~al.}}]{Sponer2016}
{\v{S}}poner, J.~E., Szabla, R., G{\'o}ra, R.~W., {et~al.} 2016, Physical
  Chemistry Chemical Physics

\bibitem[{{Stevenson}(1999)}]{Stevenson1999}
{Stevenson}, D.~J. 1999, \nat, 400, 32

\bibitem[{{Stubenrauch} {et~al.}(2013){Stubenrauch}, {Rossow}, {Kinne},
  {Ackerman}, {Cesana}, {Chepfer}, {Di Girolamo}, {Getzewich}, {Guignard},
  {Heidinger}, {Maddux}, {Menzel}, {Minnis}, {Pearl}, {Platnick}, {Poulsen},
  {Riedi}, {Sun-Mack}, {Walther}, {Winker}, {Zeng}, \&
  {Zhao}}]{Stubenrauch2013}
{Stubenrauch}, C.~J., {Rossow}, W.~B., {Kinne}, S., {et~al.} 2013, Bulletin of
  the American Meteorological Society, 94, 1031

\bibitem[{{Tarter} {et~al.}(2007){Tarter}, {Backus}, {Mancinelli}, {Aurnou},
  {Backman}, {Basri}, {Boss}, {Clarke}, {Deming}, {Doyle}, {Feigelson},
  {Freund}, {Grinspoon}, {Haberle}, {Hauck}, {Heath}, {Henry}, {Hollingsworth},
  {Joshi}, {Kilston}, {Liu}, {Meikle}, {Reid}, {Rothschild}, {Scalo}, {Segura},
  {Tang}, {Tiedje}, {Turnbull}, {Walkowicz}, {Weber}, \& {Young}}]{Tarter2007}
{Tarter}, J.~C., {Backus}, P.~R., {Mancinelli}, R.~L., {et~al.} 2007,
  Astrobiology, 7, 30

\bibitem[{{Tian}(2009)}]{Tian2009b}
{Tian}, F. 2009, \apj, 703, 905

\bibitem[{{Tian} {et~al.}(2009){Tian}, {Kasting}, \& {Solomon}}]{Tian2009a}
{Tian}, F., {Kasting}, J.~F., \& {Solomon}, S.~C. 2009, \grl, 36, L02205

\bibitem[{{Todd} {et~al.}(2017){Todd}, {Fahrenbach}, {Sasselov}, {Magnani}, \&
  {Ranjan}}]{Todd2017a}
{Todd}, Z., {Fahrenbach}, A., {Sasselov}, D., {Magnani}, C., \& {Ranjan}, S.
  2017, submitted

\bibitem[{{Toon} {et~al.}(1989){Toon}, {McKay}, {Ackerman}, \&
  {Santhanam}}]{Toon1989}
{Toon}, O.~B., {McKay}, C.~P., {Ackerman}, T.~P., \& {Santhanam}, K. 1989,
  \jgr, 94, 16287

\bibitem[{Voet {et~al.}(1963)Voet, Gratzer, Cox, \& Doty}]{Voet1963}
Voet, D., Gratzer, W.~B., Cox, R.~a., \& Doty, P. 1963, Biopolymers, 1, 193.
\newblock \url{http://doi.wiley.com/10.1002/bip.360010302}

\bibitem[{{Wordsworth} {et~al.}(2016){Wordsworth}, {Kalugina}, {Lokshtanov},
  {Vigasin}, {Ehlmann}, {Head}, {Sanders}, \& {Wang}}]{Wordsworth2017}
{Wordsworth}, R., {Kalugina}, Y., {Lokshtanov}, S., {et~al.} 2016, ArXiv
  e-prints, arXiv:1610.09697

\bibitem[{{Wordsworth} \& {Pierrehumbert}(2013)}]{Wordsworth2013h2}
{Wordsworth}, R., \& {Pierrehumbert}, R. 2013, Science, 339, 64

\bibitem[{{Wordsworth} \& {Pierrehumbert}(2014)}]{Wordsworth2014}
---. 2014, \apjl, 785, L20

\bibitem[{Wordsworth {et~al.}(2015)Wordsworth, Kerber, Pierrehumbert, Forget,
  \& Head}]{Wordsworth2015}
Wordsworth, R.~D., Kerber, L., Pierrehumbert, R.~T., Forget, F., \& Head, J.~W.
  2015, Journal of Geophysical Research E: Planets, 120, 1201

\bibitem[{Xu {et~al.}(2016)Xu, Tsanakopoulou, Magnani, Szabla, {\v{S}}poner,
  {\v{S}}poner, G{\'o}ra, \& Sutherland}]{Xu2016}
Xu, J., Tsanakopoulou, M., Magnani, C.~J., {et~al.} 2016, Nature Chemistry

\end{thebibliography}

\end{document}